\documentclass[journal=jacsat,manuscript=article]{achemso}
\usepackage[width=0.9\textwidth, font={footnotesize, singlespacing}]{subcaption}
\usepackage{amsmath}
\usepackage{amssymb}
\usepackage{mathtools}
\usepackage{bm}

\setkeys{acs}{maxauthors = 100}

\author{Nina Glaser}
\author{Alberto Baiardi}
\author{Markus Reiher}
\email{markus.reiher@phys.chem.ethz.ch}
\affiliation[ETHZ]{Laboratorium f\"ur Physikalische Chemie, ETH Z\"urich, Vladimir-Prelog-Weg 2, 8093 Z\"urich, Switzerland}

\title{Tensor Network States for Vibrational Spectroscopy}

\SectionNumbersOn

\begin{document}

\section{Introduction}

The exact solution of vibrational and electronic quantum many-body problems is obtained by solving the full configuration interaction (CI) eigenvalue equation.\cite{Shavitt1998_FCI}
However, the practical application of full CI is limited due to its exponentially growing computational cost.
This changed in the last decade, which has witnessed an impressive development of modern full CI schemes.\cite{Eriksen2020_Benzene,Williams2020_ManyBodyProblem}
In this context, methods based on so-called tensor networks states have made it possible to calculate accurate energies for systems with up to 100 single-particle basis functions.\cite{white92,White1993_DMRGBasis,Chan2008_Review,Zgid2009_Review,Schollwoeck2011_Review-DMRG,Chan2011_Review,Wouters2013_Review,Keller2014,Kurashige2014_Review,Olivares2015_DMRGInPractice,Szalay2015_Review,Yanai2015,Knecht2016_Chimia,Baiardi2020_Review}
These methods tame the exponential scaling of full CI by efficiently compressing the many-body wave function representation by tensor factorization and by optimizing the wave function with algorithms that are tailored to each specific factorization.
The potential of these methods is determined by a complex interplay between two components: representation effectiveness and availability of an efficient optimization algorithm.
The ideal tensor network state should provide an efficient description of a quantum many-body state, especially in the presence of strong correlation, where other simplified methods, based either on truncated CI or on low-order perturbation theory, fail.
Finding the optimal balance between these two components is a key challenge for the development of tensor-based methods. \\

The development of tensor-based quantum chemical methods has not followed a straight path and a variety of different algorithms have been designed for time-independent and time-dependent, electronic and vibrational many-body chemical problems.
The most popular tensor factorization for nuclear quantum dynamics is the multi-layer multi-configurational time-dependent Hartree (ML-MCTDH) algorithm.\cite{Thoss2003_ML-MCTDH,Manthe2008_MLMCTDH-Original,Vendrell2011_ML-MCTDH,Burghardt2013_ML-G-MCTDH,Wang2015_ML-MCTDH}
ML-MCTDH factorizes the full-CI tensor based on the so-called hierarchical Tucker format\cite{Lubich2013_HT-Factorization} and solves the underlying time-dependent (TD) Schr\"{o}dinger equation with the multi-layer generalization of the MCTDH method.\cite{Meyer2012_MCTDH-Review}
The ML-MCTDH accuracy relies heavily on the possibility of partitioning the quantum system into subsets of strongly-interacting particles.
It is, therefore, maximally efficient when applied to system-bath Hamiltonians, even though it has also been successfully applied to more complex molecular systems.\cite{Zie2015_MCTDH-Fullerene,Burghardt2018_Polymer-ML-MCTDH,Burghardt2019_P3HT-ML-MCTDH}
The imaginary-time formulation of ML-MCTDH enables the optimization of the ground-state of many-body Hamiltonians.\cite{Manthe2012_Malonaldehyde-ImaginaryTime} 
For vibrational calculations, Carrington and co-workers proposed the canonical polyadic factorization\cite{Leclerc2014_TensorDecomposition,Leclerc2017_RankReduced-Comparison} and its hierarchical variant\cite{Thomas2015_NestedContractions} to represent anharmonic vibrational wave functions for molecules with more than 30 atoms.
The same factorization has also been combined with other wave function ans\"{a}tze that do not rely on the full-CI parameterization.
For instance, Christiansen and co-workers applied tensor decomposition algorithms to encode compactly vibrational coupled cluster wave functions.\cite{godtliebsen13,godtliebsen15} \\
The development of tensor network states for electronic-structure calculations has focused on a third class of tensor factorizations.
The reference method for electronic problems has been the density matrix renormalization group (DMRG) algorithm\cite{white92,White1993_DMRGBasis} which relies on the so-called tensor train (TT) factorization.\cite{Oseledets2012_ALS}
It is, in fact, possible to demonstrate that the TT factorization, also known as matrix product state (MPS) representation in physics,\cite{McCulloch2007_FromMPStoDMRG} efficiently encodes the ground-state wave function of short-ranged Hamiltonians.\cite{Hastings2007_AreaLaw}
The success of DMRG is not only rooted in the fact that for such systems there are well-defined conditions under which its efficiency is ensured.
Originally devised for one-dimensional spin chains in condensed matter physics, DMRG has later been applied to molecular electronic structure problems.\cite{Shuai1996_Original,fano98,Shuai1998_PPP,white99,chan02,Legeza2003_DMRG-LiF}
It turned out that DMRG is also an efficient strategy to approximate states that have no inherently (pseudo-)one-dimensional structure, such as strongly correlated systems in transition metal complexes and clusters.\cite{marti08}
Since then, it has become one of the reference method for large-scale CI electronic-structure calculation.\cite{Legeza2008_Review,Chan2008_Review,Zgid2009_Review,Marti2010_Review-DMRG,Schollwoeck2011_Review-DMRG,Chan2011_Review,Wouters2013_Review,Kurashige2014_Review,Olivares2015_DMRGInPractice,Szalay2015_Review,Yanai2015,Knecht2016_Chimia,Baiardi2020_Review}\\

DMRG has recently been generalized to vibrational problems and was applied to anharmonic vibrational-structure calculations.\cite{Oseledts2016_VDMRG,baiardi17}
The resulting theory, abbreviated vDMRG, delivers accurate anharmonic energies for molecules with up to 30 atoms, i.e. for system sizes that are the current limits of alternative tensor-based methods.
Moreover, methods originally designed for large-scale electronic full CI calculation have been applied to vibrational problems, including multidimensional DMRG generalizations\cite{larsson19} and selected CI\cite{Brorsen2020_PreBO-HBCI,Berkelbach2021_HBCI-Vibrational} algorithms, which are as accurate as traditional state-of-the-art vibrational full CI algorithms.
This emphasizes that tensor factorizations for vibrational problems can further expand the range of applicability of anharmonic vibrational-structure calculations.\\

DMRG is, in fact, the most popular tensor-based method for quantum-chemical applications due to its numerical stability and its flexibility, which makes it applicable to many different quantum chemical problems.
Moreover, it represents the foundation of more complex tensor-based algorithms.
In this chapter, we introduce the tensor network state framework with a particular focus on the DMRG algorithm.
We will first review the theoretical foundations of the DMRG.
Then, we discuss the optimization of ground- and excited-state anharmonic vibrational wave functions with DMRG.
Afterwards, we elaborate on extensions of the DMRG to solve the nuclear time-dependent Schr\"{o}dinger equation, leading to the time-dependent DMRG (TD-DMRG) method.
Finally, we explain the main shortcomings and limitations of the current vDMRG variants, and discuss possible strategies to overcome them.

\section{Tensor Decompositions and Tensor Network States}

The description of quantum many-body systems with tensor networks has become increasingly important in recent years.\cite{orus19}
Tensor decomposition can be understood as a fragmentation scheme for factorizing the wave function into smaller pieces.
The parametrization of the many-body wave function is decomposed into a multitude of tensors, which are connected according to a specific pattern reflecting the entanglement structure of the system.
The resulting wave function representation is known as a tensor network state.\\

\subsection{Diagrammatic Notation of Tensor Networks}

\begin{figure}[ht]
  \centering
     \begin{subfigure}[t]{0.12\textwidth}
       \centering
       \includegraphics[height=0.4cm]{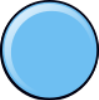}
       \caption{Scalar}
     \end{subfigure}
  \hspace{0.5cm}
  \centering
     \begin{subfigure}[t]{0.12\textwidth}
         \centering
         \includegraphics[height=0.4cm]{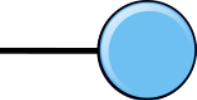}
         \caption{Vector}
     \end{subfigure}
  \hspace{0.5cm}
  \centering
     \begin{subfigure}[t]{0.12\textwidth}
         \centering
         \includegraphics[height=0.4cm]{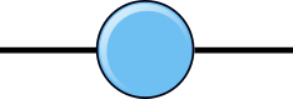}
         \caption{Matrix}
     \end{subfigure}
  \hspace{0.5cm}
  \centering
     \begin{subfigure}[t]{0.20\textwidth}
         \centering
         \includegraphics[height=0.8cm]{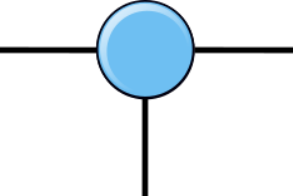}
         \caption{Tensor of rank 3}
     \end{subfigure}
     \caption{Diagrammatic notation of tensors up to rank 3}
     \label{fig:tensors}
\end{figure}

Tensors are multidimensional arrays of (real or complex) numbers, with their rank corresponding to the number of indices required to specify their elements.
Hence, a tensor of rank zero is a scalar, rank one corresponds to a vector, rank two denotes a matrix, and so forth.
Tensors are conveniently represented with a graphical notation as introduced in Ref.~\citenum{penrose71}, where the tensor core is represented by a solid shape, and its indices are depicted by lines, as shown in Fig.~\ref{fig:tensors}.

\begin{figure}[ht]
	\centering
     \begin{subfigure}[t]{0.25\textwidth}
         \centering
         \includegraphics[height=0.4cm]{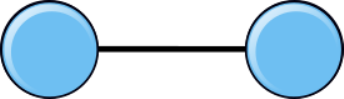}
         \caption{Scalar product of vectors}
     \end{subfigure}
     \hspace{0.5cm}
     \centering
     \begin{subfigure}[t]{0.2\textwidth}
         \centering
         \includegraphics[height=0.4cm]{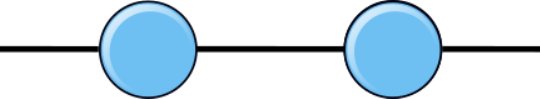}
         \caption{Matrix product}
     \end{subfigure}
     \hspace{0.5cm}
     \centering
     \begin{subfigure}[t]{0.3\textwidth}
         \centering
         \includegraphics[height=1.2cm]{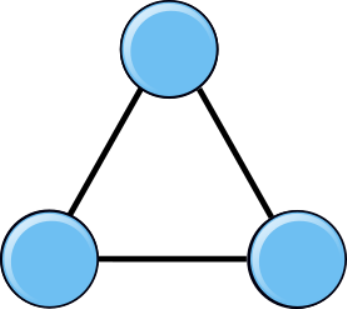}
         \caption{Trace of product of 3 matrices}
     \end{subfigure}
     \caption{Tensor network diagrams of different index contractions}
     \label{fig:tns_dia}
\end{figure}

A core attribute of tensors is that high-rank tensors can be decomposed into a multitude of lower-rank tensors connected in a network.
Mathematically, such a tensor network is a set of tensors where the indices are contracted according to a particular pattern, with index contraction referring to summing over all possible values of the repeated indices of a set of tensors.
Contractions are graphically represented by joining index lines in tensor network diagrams, as illustrated in Fig.~\ref{fig:tns_dia}.
A variety of different tensor network topologies can be realized when decomposing a large-rank tensor.
Ideally, the chosen network should provide a very compact representation of the original tensor.

\subsection{Tensor Networks for Quantum States}
\begin{figure}[ht]
  \centering
  \begin{subfigure}[t]{0.4\textwidth}
    \centering
    \includegraphics[height=0.8cm]{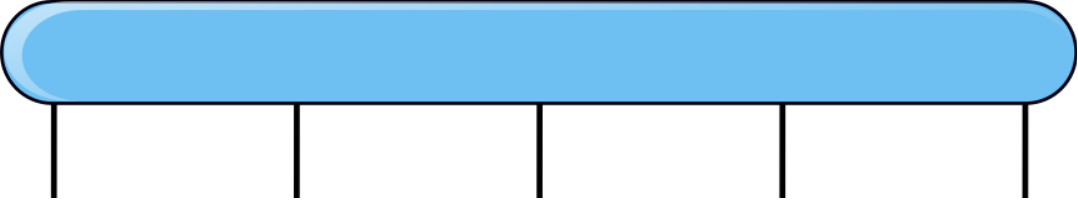}
    \caption{CI tensor $\textbf{C}$}
    \label{fig:citensor}
  \end{subfigure}
  \hspace{0.5cm}
  \begin{subfigure}[t]{0.3\textwidth}
    \centering
    \includegraphics[height=0.8cm]{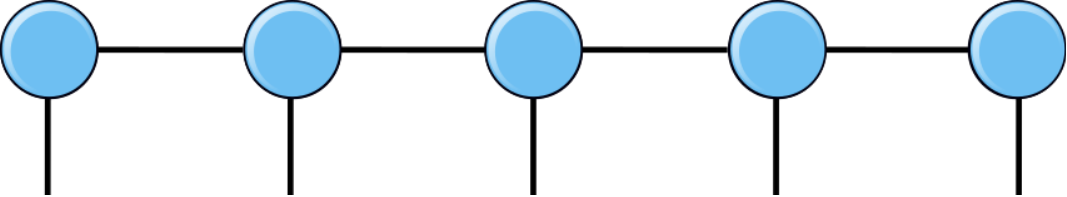}
    \caption{Matrix product state}
    \label{fig:mps}
  \end{subfigure}
  \hfill
  \caption{Tensor diagrams of the CI coefficient tensor (left panel) and of the corresponding matrix product state (right panel) for a 5-site system.
  The open indices in both diagrams correspond to the physical degrees of freedom of the local Hilbert spaces spanned by $\vert \sigma_i \rangle$.}
\end{figure}

Tensor networks can be employed to parametrize quantum many-body states.
An arbitrary many-body system with $L$ basis functions is described by the full CI wave function, which can be written as

\begin{eqnarray}
  \vert \Psi \rangle 
    &=& \sum_{\mathclap{\sigma_1, \sigma_2, ..., \sigma_L}} c_{\sigma_1, \sigma_2, ..., \sigma_L}
        \vert \sigma_1 \rangle  \otimes \vert \sigma_2 \rangle  \otimes ...  \otimes \vert \sigma_L \rangle 
  \label{eq:fci_1} \\
    &=& \sum_{\boldsymbol{\sigma}} c_{\boldsymbol{\sigma}} \vert \boldsymbol{\sigma} \rangle \, ,
  \label{eq:fci}
\end{eqnarray}
where $\vert \sigma_i \rangle$ represents the possible occupations of the $i$-th basis function, $\otimes$ denotes the tensor product, and $\vert \boldsymbol{\sigma} \rangle$ is the occupation number vector.
The CI coefficients $c_{\boldsymbol{\sigma}}$ can be collected in one large tensor $\textbf{C}$ of rank $L$, as visualized in Fig.~\ref{fig:citensor}.
A compact representation of the tensor $\textbf{C}$ is obtained by replacing it with a network of smaller tensors.
For this purpose, the chosen tensor network should be flexible enough to capture the relevant correlation and its topology should reflect the entanglement structure of the underlying many-body wave function.
We review four of the most common tensor network types in quantum chemistry, each of them corresponding to a different factorization of the CI tensor.
We start with the matrix product state (MPS), and demonstrate how the full CI wave function can be decomposed into an MPS, before briefly introducing tree tensor network states (TTNS), projected entangled pair states (PEPS), and complete graph tensor network states (CGTNS).\\

One of the simplest and most common tensor decompositions of a many-body wave function is the tensor train (TT) factorization\cite{kolda09}, which encodes the wave function as an MPS.
In an MPS, the rank-$L$ full-CI tensor $\textbf{C}$ is factorized as the product of $L$ three dimensional tensors, one per single-particle basis function of the many-body system,
\begin{eqnarray}
  \vert \Psi \rangle &=& \sum\limits_{\sigma_1 \ldots \sigma_L} \sum\limits_{a_1 ... a_{L-1 }} 
    M_ {1 a_1}^{\sigma_1} M_ {a_1 a_2}^{\sigma_2} \cdots M_ {a_{L-1} 1}^{\sigma_L} 
   \vert \sigma_1 \rangle  \otimes \vert \sigma_2 \rangle  \otimes ...  \otimes \vert \sigma_L \rangle 
   \label{eq:MPS_1} \\
  &=& \sum\limits_{\boldsymbol{\sigma}} \textbf{M}^{\sigma_1} \textbf{M}^{\sigma_2} \cdots 
                                       \textbf{M}^{\sigma_L}  \vert \boldsymbol{\sigma} \rangle \, ,
  \label{eq:MPS_2}
\end{eqnarray}
where the CI coefficients $c_{\boldsymbol{\sigma}}$ are obtained as the product of a set of matrices $\textbf{M}^{\sigma_l}$.
The diagrammatic notation of an MPS is given in Fig.~\ref{fig:mps}.

\begin{figure}[ht]
  \centering
  \includegraphics[width=4.0cm]{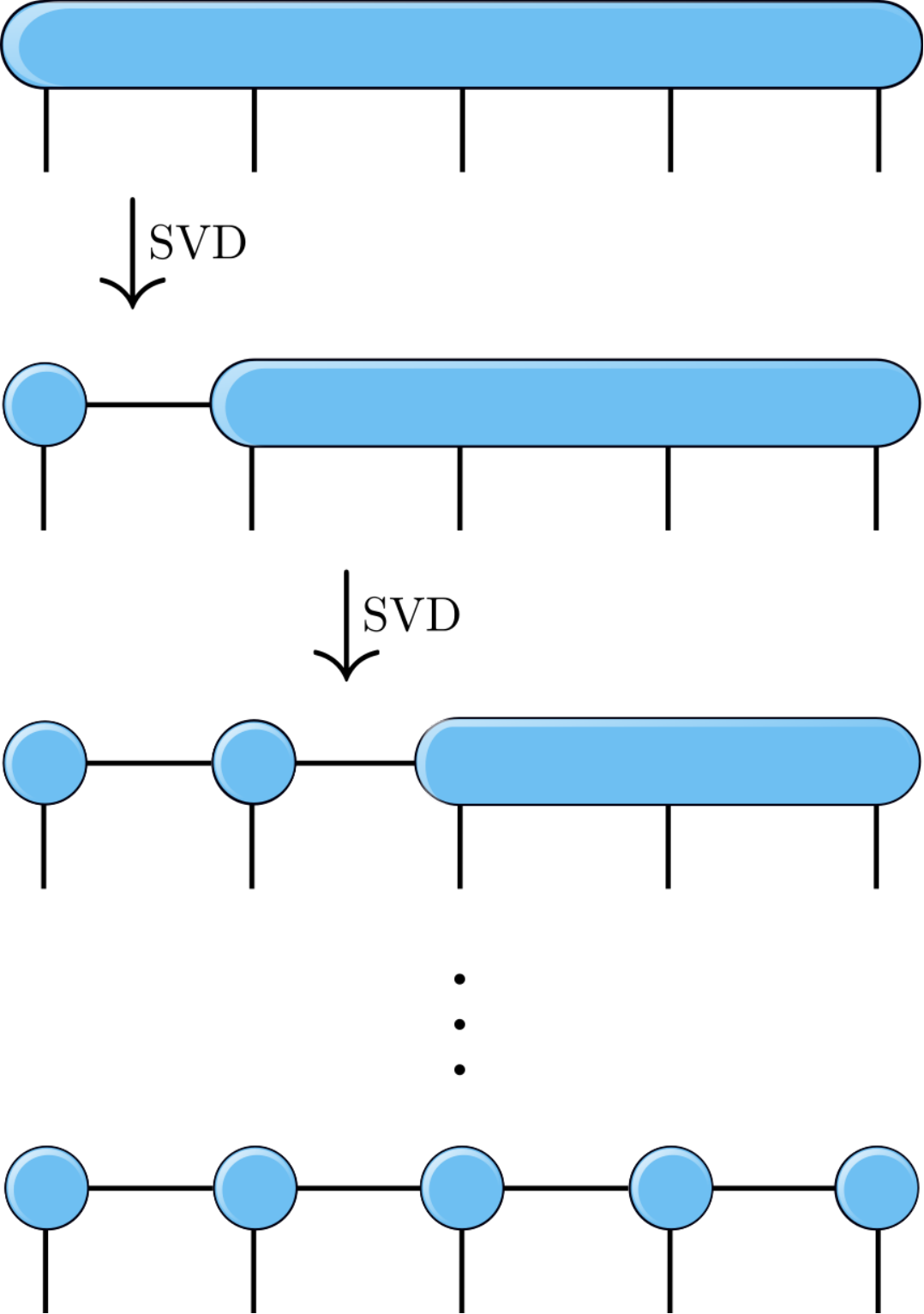}
  \caption{Graphical representation of the decomposition of the CI tensor into an MPS by a sequence of singular value decompositions.}
  \label{fig:decomp}
\end{figure}

The exact MPS representation of a full CI wave function is constructed by reshaping the CI tensor $\textbf{C}$ of rank $L$ into a matrix $\textbf{M}_{\sigma_1, (\sigma_2 \dots \sigma_L)}$ with dimension $N \times N^{L-1}$, where $N$ is the number of possible single-particle states, and then calculating the singular value decomposition (SVD) of the resulting matrix.
Any general $p \times n$ matrix $\textbf{M}$ can be factorized as

\begin{equation}
  \textbf{M} = \textbf{U} \textbf{S} \textbf{V}^{\dagger} \, ,
  \label{eq:SVD}
\end{equation}
where matrices \textbf{U} and \textbf{V} have orthogonal columns, and \textbf{S} is a diagonal matrix with real non-negative elements, the singular values.
The dimensions of the decomposed matrices are $p \times \min (p,n)$ for \textbf{U}, $\min (p,n) \times n$ for \textbf{V}, and $\min (p,n) \times \min (p,n)$ for \textbf{S}.
By repeatedly applying SVD to the reshaped matrices of the CI coefficients, as illustrated in Fig.~\ref{fig:decomp}, the CI wave function can be decomposed into a product of $L$ matrices, hence resulting in an MPS parametrization.
The linear arrangement of the tensors is referred to as the MPS lattice, with each tensor being attributed to a specific site on the lattice.

The exact decomposition of a full CI state results in an MPS with matrix dimensions $1 \times N$, $N \times N^2$, $\dots$, $N^{\frac{L}{2} -1} \times N^{\frac{L}{2}}$, $N^{\frac{L}{2}} \times N^{\frac{L}{2} -1}$, $\dots$, $N^2 \times N$, $N \times 1$ for $\textbf{M}^{\sigma_1}$ to $\textbf{M}^{\sigma_L}$, respectively.
Hence, the number of entries still scales exponentially with $L$, as for the original CI tensor.
The key advantage of the MPS representation is that each SVD step (see Eq.~(\ref{eq:SVD})) can be truncated to control the size of the factorization.
This is achieved by limiting the dimension of the matrices $\textbf{M}^{\sigma_i}$ to some maximum size of $m \times m$, where $m$ is referred to as the bond dimension.
In practice, this truncation step is realized by setting all but the $m$ largest diagonal elements in \textbf{S} to zero, and retaining only the $m$ first columns of \textbf{U} and \textbf{V}.
The error introduced in each truncation step can be straightforwardly evaluated based on the sum of the discarded singular values.\cite{legeza03}\\

The bond dimension $m$ is the key parameter of the MPS wave function representation.
A maximal reduction of the bond dimension to $m$=1 results in an ordinary product state, which corresponds to the mean-field reference that neglects all correlation within the system.
The MPS representation can be enhanced by increasing the bond dimension, with more correlated systems requiring higher values of $m$ to accurately encode many-particle correlation effects.
The resulting MPS wave function of bond dimension $m$ scales polynomially, while also accounting for correlation between the different sites, therefore providing a compact parametrization of the full CI problem.
The efficiency of the MPS compression is ensured by the area law,\cite{Hastings2007_AreaLaw} which entails that for short-ranged one-dimensional Hamiltonians the bond dimension $m$ required to represent the ground state with a given accuracy is independent of the system size $L$, therefore taming the exponential scaling of the FCI wave function.
While most Hamiltonians encountered in quantum chemistry do not satisfy the area law conditions, the wave functions can nevertheless be efficiently parametrized as MPS.
It has been shown that even though the MPS truncation introduces an approximation of the FCI wave function for long-range Hamiltonians, MPS with significantly reduced bond dimensions can accurately represent a broad range of systems.\cite{Legeza2008_Review,Chan2008_Review,marti08,Zgid2009_Review,Marti2010_Review-DMRG,Schollwoeck2011_Review-DMRG,Chan2011_Review,Wouters2013_Review,Kurashige2014_Review,Olivares2015_DMRGInPractice,Szalay2015_Review,Yanai2015,Knecht2016_Chimia,Baiardi2020_Review}
Hence, the MPS decomposition provides one of the most powerful representations of molecular wave functions in the presence of strong correlation effects. \\

\begin{figure}[ht]
  \centering
  \begin{subfigure}[t]{0.25\textwidth}
    \centering
    \includegraphics[height=2.7cm]{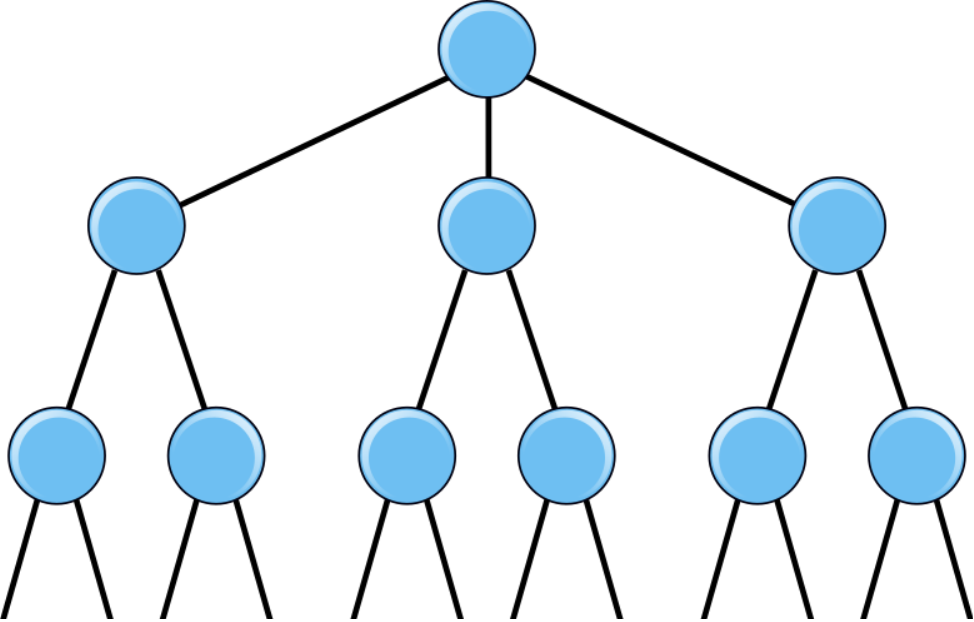}
    \caption{Tree tensor network state (TTNS)}
  \end{subfigure}
  \hspace{0.5cm}
  \centering
  \begin{subfigure}[t]{0.3\textwidth}
    \centering
    \includegraphics[height=3.0cm]{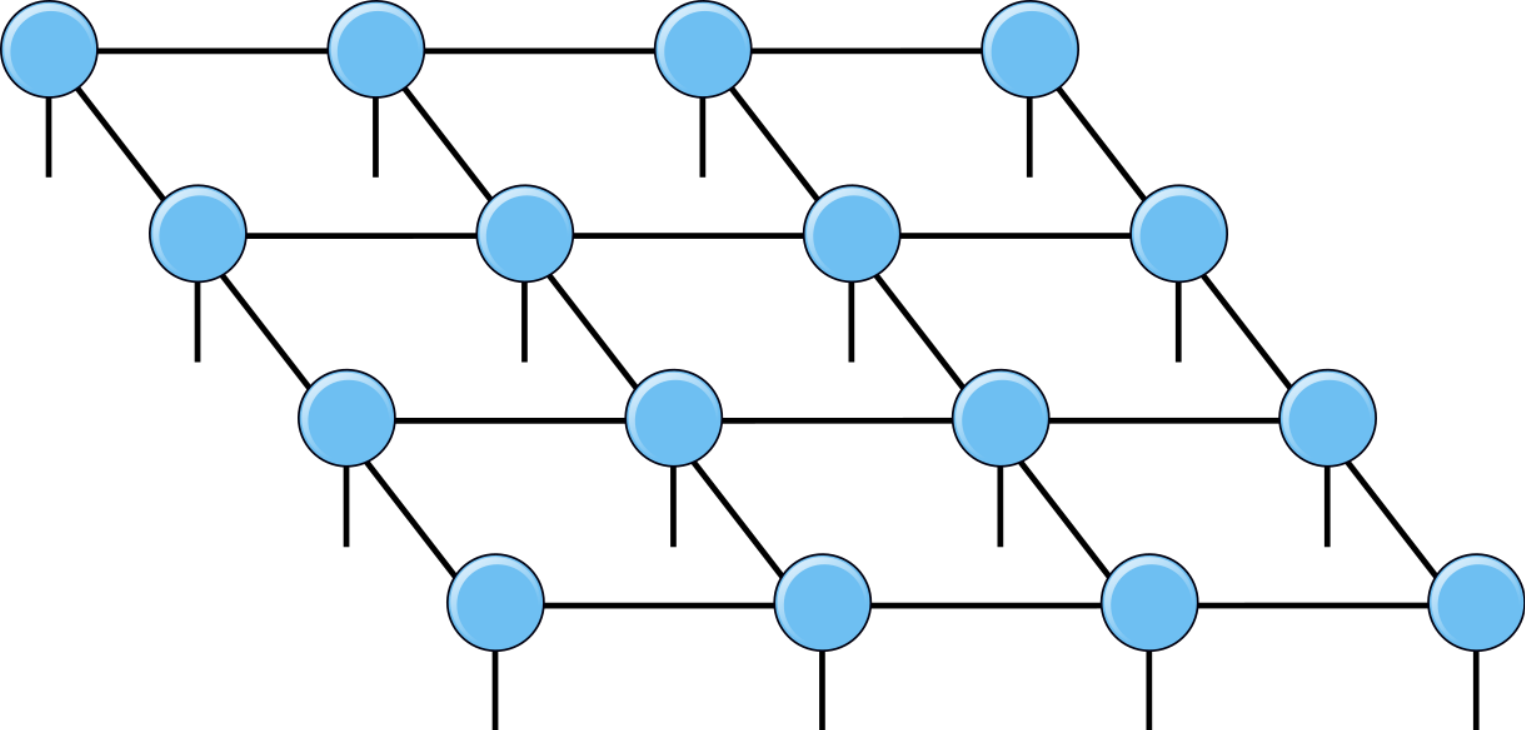}
    \caption{Projected entangled pair state (PEPS)}
  \end{subfigure}
    \hspace{1.5cm}
  \begin{subfigure}[t]{0.25\textwidth}
    \centering
    \includegraphics[height=3.5cm]{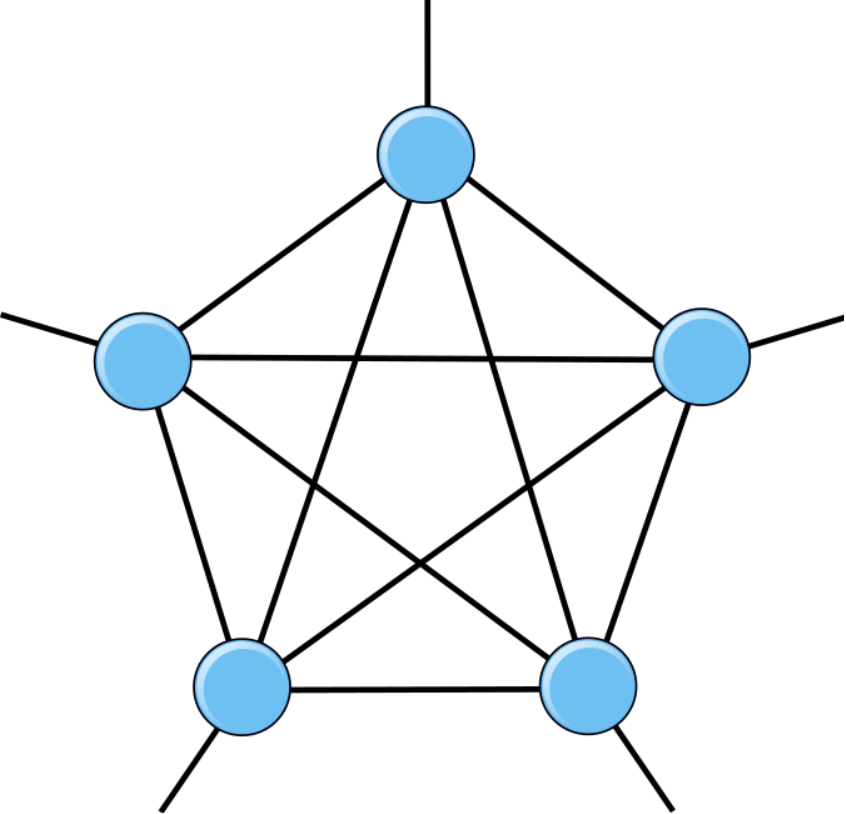}
    \caption{Complete graph tensor network state (CGTNS)}
  \end{subfigure}
  \hfill
  \caption{Tensor network diagrams of MPS generalizations}
  \label{fig:ttns_peps}
\end{figure}

While MPS yield extremely compact wave function representations for quasi monodimensional Hamiltonians, other tensor network geometries might encode the entanglement structure of certain systems more efficiently.
Two of the simplest generalizations of MPS are tree tensor network states (TTNS)\cite{Noack2010_TTNS,nakatani13,Legeza2015_TTNS,Gunst2018_3LTNS,larsson19} and projected entangled pair states (PEPS).\cite{Verstraete2010_PEPS,Chan2019_PEPS}
While matrix product states encode a one-dimensional correlation structure, TTNS employ tree-like connections, whereas PEPS generalize the MPS structure to two spatial dimensions.
Another commonly employed tensor factorization in quantum chemistry is the complete graph tensor network state (CGTNS),\cite{Marti_2010,Kovyrshin_2016,kovyrshin17} where each site is connected to every other site, as illustrated in Figure~\ref{fig:ttns_peps}.

Compared to MPS, TTNS allow for a more flexible representation of the system structure, 
whereas PEPS are able to directly encode correlation in two dimensional quantum systems and CGTNS account for all interactions between any pair of sites.
While these geometries can lead to more compact wave function parametrizations, more complex connection graphs also result in higher computational costs for optimizing their parameters.
Hence, there is often a dilemma between tensor network state flexibility and computational scaling, because the numerical optimization of the tensor network becomes less efficient for more complex tensor factorizations. \\

One of the most commonly applied tensor network based algorithms is the density matrix renormalization group (DMRG), which is based on an efficient deterministic variational optimization of matrix product states.
Tensor network states with a more complex structure than MPS are often variationally optimized with Monte Carlo-based optimization algorithms.\cite{Vidal2007_VariationalQMC}
The development of increasingly efficient sampling schemes\cite{Sharma2018_OrbitalSpaceVMC} is constantly enhancing the efficiency of these stochastic optimization algorithms.\cite{Marti_2010,kovyrshin17}
However, they all share a common limitation, namely that the wave function is optimized in the full CI space.
Therefore, they do not fully exploit the compact structure of the underlying tensor factorization.
A notable exception are TTNSs, for which a DMRG-like optimization strategy has been devised.\cite{nakatani13}
Moreover, the recently developed time-dependent TTNS variant\cite{Kloss2020_TD-TTNS,Lubich2021_TTNS-TD} has paved the route towards the optimization of TTNS with imaginary-time propagation algorithms.
Although these developments could set the ground for new efficient tensor-based methods in the near future, DMRG currently provides the optimal balance between the complexity of the tensor factorization and of its optimization.
Therefore, we will focus on DMRG-based algorithms for the remainder of this chapter.

\section{MPS/MPO Formulation of the Density Matrix Renormalization Group}

Finding the optimal tensor network state representation of a quantum many-body wave function is a very challenging task that requires solving complex non-linear optimization problems.
For wave functions encoded as MPS, an efficient optimization technique is offered by the density matrix renormalization group (DMRG) algorithm.
DMRG was first introduced in solid state physics in 1992 by White\cite{white92} as a method to calculate ground-state wave functions of one-dimensional quantum lattices.
In its original formulation, DMRG was not contrived as a tensor network approach, but relied on the theoretical framework of the numerical renormalization group.\cite{fisher74}
The tensor network-based formulation of DMRG was developed later, when it was discovered that the entire problem can be reformulated in terms of matrix product parametrizations.\cite{ostlund95,McCulloch2007_FromMPStoDMRG}
This second formulation proved especially useful for quantum chemical applications, as it straightforwardly supports arbitrarily complex Hamiltonians.
Therefore, it will be the main focus for the remainder of this chapter.

\subsection{Variational Optimization in the MPS/MPO Framework}

As a variational method, DMRG optimizes the MPS parameters by minimizing the expectation value of the Hamiltonian $\langle \Psi \vert \hat{\mathcal{H}} \vert \Psi \rangle$ for normalized wave functions $\Psi$.
In practice, the Lagrangian $\mathcal{L}$
\begin{equation}
  \mathcal{L} = \langle \Psi \vert \hat{\mathcal{H}} \vert \Psi \rangle - \lambda \left( \langle \Psi \vert \Psi \rangle - 1 \right) \, 
  \label{eq:lagrangian}
\end{equation}
is minimized, with $\lambda$ being the Lagrangian multiplier enforcing the normalization constraint.
The expectation value of the Hamiltonian is conveniently minimized by expressing the Hamiltonian operator in a form analogous to the MPS, known as a matrix product operator (MPO).
A general operator $\mathcal{\hat{W}}$ of the form

\begin{equation}
  \hat{\mathcal{W}} = \sum\limits_{\boldsymbol{\sigma}, \boldsymbol{\sigma}'} w_{\boldsymbol{\sigma}, \boldsymbol{\sigma}'} \vert \boldsymbol{\sigma} \rangle \langle \boldsymbol{\sigma}' \vert
  \label{eq:GenericOperator}
\end{equation}
can be exactly expressed as an MPO

\begin{equation}
  \mathcal{\hat{W}} = \sum\limits_{\boldsymbol{\sigma}, \boldsymbol{\sigma}'} \sum\limits_{b_1 \dots b_{L-1}}
                      W_{1 b_1}^{\sigma_1 \sigma_1'} W_{b_1 b_2}^{\sigma_2 \sigma_2'} \cdots W_{b_{L-1} 1}^{\sigma_L \sigma_L'} \vert
                      \boldsymbol{\sigma} \rangle \langle \boldsymbol{\sigma}' \vert \, .
  \label{eq:MPO}
\end{equation}

By introducing

\begin{equation}
  \hat{W}_{b_{l-1} b_l}^{[l]} = \sum\limits_{\sigma_l, \sigma_l'} W_{b_{l-1} b_l}^{\sigma_l \sigma_l'} \vert \boldsymbol{\sigma} \rangle \langle \boldsymbol{\sigma}' \vert
  \label{eq:MPOTensor}
\end{equation}
we can simplify the MPO expression to 

\begin{eqnarray}
  \mathcal{\hat{W}} &=& \sum_{\mathclap{{b_1 \dots b_{L-1}}}} \hat{W}_{1 b_1}^{[1]} \hat{W}_{b_1 b_2}^{[2]} \cdots \hat{W}_{b_{L-1} 1}^{[L]} \vert \boldsymbol{\sigma} \rangle \langle \boldsymbol{\sigma}' \vert \\
&=& \hat{\textbf{\text{W}}}^{[1]} \hat{\textbf{\text{W}}}^{[2]} \cdots \hat{\textbf{\text{W}}}^{[L]} \, .
\end{eqnarray}

\begin{figure}[ht]
  \centering
  \begin{subfigure}[t]{0.4\textwidth}
    \centering
    \includegraphics[height=1.5cm]{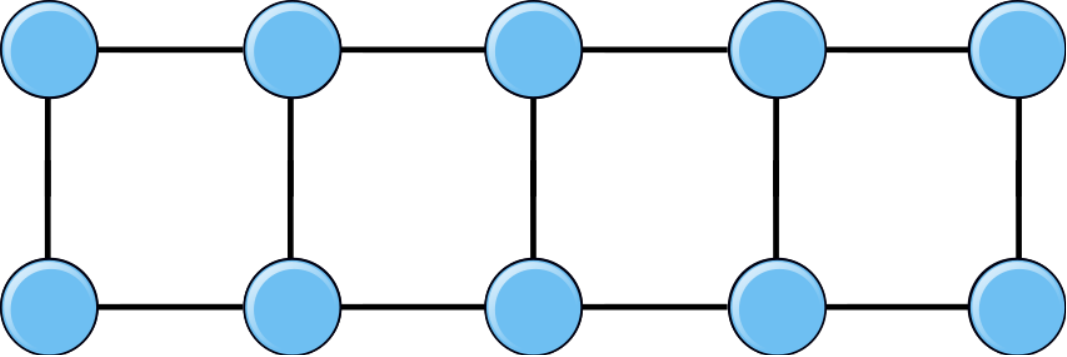}
    \caption{Overlap $\langle \Psi \vert \Psi \rangle$}
  \end{subfigure}
  \hspace{1cm}
  \centering
  \begin{subfigure}[t]{0.4\textwidth}
    \centering
    \includegraphics[height=2.0cm]{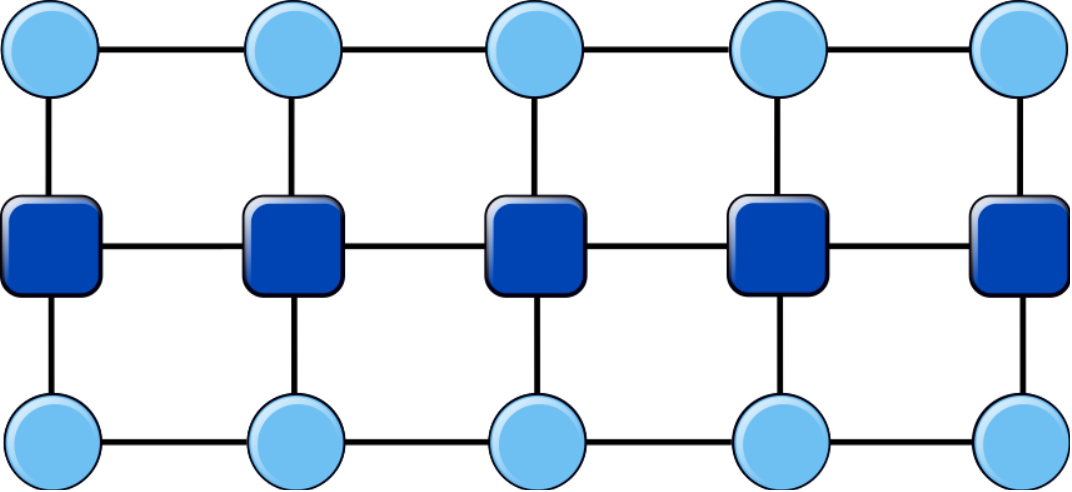}
    \caption{Expectation value $\langle \Psi \vert \hat{W} \vert \Psi \rangle$}
    \label{fig:mps_mpo}
  \end{subfigure}
  \hfill
  \caption{Tensor network diagrams representing the calculation of the overlap between two MPSs (a) and the expectation values of the Hamiltonian, expressed as an MPO, over an MPS (b). In the right panel, the dark blue squares represent the matrix product operator $\hat{W}$, while the light blue circles denote the MPS tensors of the wave function $\Psi$.}
  \label{fig:expectvalues}
\end{figure}

The matrices $\hat{\textbf{\text{W}}}^{[l]}$ are operator-valued matrices that collect the elementary operators acting on site $l$.
In the MPS/MPO framework, the calculation of both the overlap and the expectation value can be interpreted in terms of index contractions of tensor networks.
The corresponding tensor network diagrams are shown in Fig.~\ref{fig:expectvalues}.

The bond dimensions of an MPO are uniquely determined by the definition of the operator in the given basis set.
As discussed, for instance, in Ref.~\citenum{keller15}, the maximum value of the MPO bond dimension increases with the number of long-range interaction terms.
For example, for two-body operators, the MPO bond dimension grows as $\mathcal{O}(L^3)$, and for three-body operators it grows as $\mathcal{O}(L^5)$.
Larger MPO bond dimensions not only increase the memory requirements of a calculation, but also the computational cost of contracting the MPO with the MPS to evaluate the expectation value.
The calculation time can be decreased by compressing the MPO.
Analogous to the MPS compression, the MPO bond dimension can be reduced by successively applying truncated SVDs up to a given target accuracy.\cite{Dur2010_MPOConstruction,Hubig2017_MPOGeneric}

\subsection{Canonical Form}

While the evaluation of the Lagrangian is, in principle, possible for any form of the MPS, 
the corresponding equations are expressed in a much more convenient form if the MPS is brought into the so-called canonical form.
The canonical transformation lays the foundation for an efficient optimization with the DMRG algorithm, as described in the Section~\ref{subsec:opt}. \\

The MPS representation of a given wave function is not unique, because an identity matrix $\textbf{I} = \textbf{XX}^{-1}$ can be inserted between any two matrices $\textbf{M}^{\sigma_l}$ and $\textbf{M}^{\sigma_{l+1}}$ on the MPS lattice.
Hence, the overall MPS will not change if the matrix $\textbf{M}^{\sigma_l}$ is replaced by $\textbf{M}^{\sigma_l}\textbf{X}$ while $\textbf{M}^{\sigma_{l+1}}$ is replaced by $\textbf{X}^{-1}\textbf{M}^{\sigma_{l+1}}$.
By virtue of this gauge freedom, the MPS can be brought into the canonical form.
In the so-called left-canonical form

\begin{equation}
  \vert \Psi \rangle = \sum\limits_{\boldsymbol{\sigma}} \textbf{A}^{\sigma_1} \textbf{A}^{\sigma_2} \cdots \textbf{A}^{\sigma_L}  \vert \boldsymbol{\sigma} \rangle
  \label{fig:LeftCanonical}
\end{equation}
all matrices $\textbf{A}^{\sigma_l}$ are left-normalized, meaning that

\begin{equation}
  \sum\limits_{i=1}^{N} \textbf{A}^{\sigma_i \dagger} \textbf{A}^{\sigma_i} = \textbf{I} \, ,
  \label{eq:LeftNormalization}
\end{equation}
where the sum runs over the local basis at the site $l$.
Alternatively, the MPS can be brought into the right-canonical form where

\begin{equation}
  \vert \Psi \rangle = \sum\limits_{\boldsymbol{\sigma}} \textbf{B}^{\sigma_1} \textbf{B}^{\sigma_2} \cdots \textbf{B}^{\sigma_L}  \vert \boldsymbol{\sigma} \rangle \, ,
  \label{eq:RightCanonical}
\end{equation}
with right-normalized matrices $\textbf{B}^{\sigma_l}$ such that

\begin{equation}
  \sum\limits_{i=1}^{N} \textbf{B}^{\sigma_i} \textbf{B}^{\sigma_i \dagger} = \textbf{I} \, .
  \label{eq:RightNormalization}
\end{equation}

The tensor associated with site $l$, $\textbf{M}^{\sigma_l}$ with elements $M_{a_{l-1}a_l}^{\sigma_l}$, can be left-normalized by reshaping it into a single matrix with elements $M_{\left(\sigma_l a_{l-1} \right), a_l}$.
SVD is then applied to the resulting matrix with

\begin{equation}
  M_{\left(\sigma_l a_{l-1} \right), a_l} = \sum\limits_{s_l} U_{\left(\sigma_l a_{l-1} \right), s_l} S_{s_l, s_l} V_{s_l, a_l}^\dagger \, .
  \label{eq:SVD_Normalization}
\end{equation}

The matrix $U_{\left(\sigma_l a_{l-1} \right), s_l}$ can be reshaped back into a tensor with elements $A_{a_{l-1} s_l}^{\sigma_l}$  obeying the left-normalization condition.
The remaining components of the SVD, i.e. $\textbf{SV}^\dagger$, are multiplied into the matrices at the next site to update the tensor $\textbf{M}^{\sigma_{l+1}}$.
The left-canonical form of an MPS is obtained with sequential SVDs starting from the left, until the end of the lattice is reached.
Equivalently, the right-canonical form is obtained with sequential SVDs starting from the right, at each step replacing the tensor $\textbf{M}^{\sigma_l}$ by the reshaped matrix $\textbf{V}^\dagger$ and updating the tensor to the left with the remainder of the SVD by multiplying $\textbf{US}$ into the matrices at site $l-1$. \\

There exists also a so-called mixed-canonical form\cite{Schollwoeck2011_Review-DMRG,Freitag2020_BookChapter}
\begin{equation}
  \vert \Psi \rangle = \sum\limits_{\boldsymbol{\sigma}} \textbf{A}^{\sigma_1} \cdots \textbf{A}^{\sigma_{l-1}} \textbf{M}^{\sigma_l} \textbf{B}^{\sigma_{l+1}} \cdots \textbf{B}^{\sigma_L}  \vert \boldsymbol{\sigma} \rangle \, ,
  \label{eq:MixedCanonicalForm}
\end{equation}
where all matrices to the left of an arbitrarily chosen site $l$ are left-normalized, while all remaining matrices to the right are right-normalized.
This form is of special importance in DMRG, because it facilitates the efficient sweep-based optimization of the MPS, which will be introduced in Section \ref{subsec:opt} below.

\begin{figure}[ht]
  \centering
  \includegraphics[height=0.8cm]{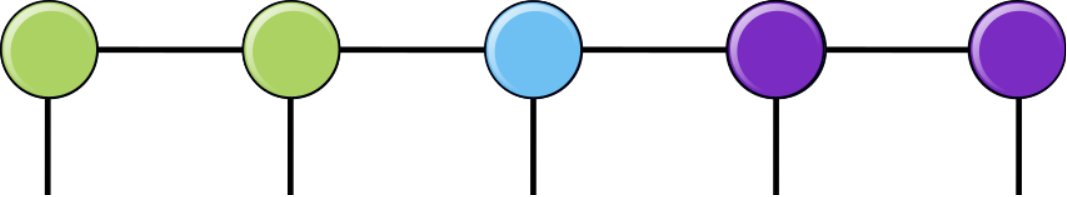}
  \caption{5-site MPS in mixed canonical form. The light green circles represent left-normalized MPS tensors $\textbf{A}$, while the purple ones denote right-normalized tensors $\textbf{B}$. The light blue circle in the center represents an unnormalized MPS tensor $ \textbf{M}$.}
  \label{fig:mps_canonical}
\end{figure}

\subsection{Expectation Values}
\label{subsec:expval}

A general expectation value of an operator in MPO form over an MPS can be written as\cite{Schollwoeck2011_Review-DMRG,keller15,Freitag2020_BookChapter}

\begin{equation}
  \langle \Psi \vert \hat{\mathcal{W}} \vert \Bar{\Psi} \rangle 
    = \sum_{\mathclap{{\substack{a_1', \dots , a_{L-1}' \\ a_1, \dots , a_{L-1} \\ \sigma \sigma'}}}} \left( M_{1 a_1}^{\sigma_1} \cdots M_{a_{L-1} 1}^{\sigma_L} \right)^{\ast}
      \sum\limits_{\mathclap{{b_1, \dots , b_{L-1}}}} W_{1 b_1}^{\sigma_1 \sigma_1'} \cdots W_{b_{L-1} 1}^{\sigma_L \sigma_L'} \left( \Bar{M}_{1 a_1'}^{\sigma_1'} \cdots \Bar{M}_{a_{L-1}' 1}^{\sigma_L'} \right) \, .
  \label{eq:expectation_no_boundaries}
\end{equation}

The order in which the sums in Eq.~(\ref{eq:expectation_no_boundaries}) are evaluated determines the computational cost of the expectation value evaluation.
For instance, if all sums over the $a$ and $b$ indices are calculated first, and then the sum over the $\sigma$ indices is evaluated, the wave function and operators will first be converted into the CI format, and only afterwards the expectation value will be calculated.
In this way, the compact form of the MPS wave function is not exploited and nothing has been gained.
By contrast, it is highly efficient to regroup the terms entering Eq.~(\ref{eq:expectation_no_boundaries}) as

\begin{equation}
  \begin{split}
    &\langle \Psi \vert \hat{\mathcal{W}} \vert \Bar{\Psi} \rangle = \\
    &\sum_{\mathclap{\substack{a_{L-1} a_{L-1}' b_{L-1} \\ \sigma_L \sigma_L'}}}  M_{a_{L-1} 1}^{\sigma_L \ast} W_{b_{L-1} 1}^{\sigma_L \sigma_L'} \left( \cdots \left( 
       \sum_{\substack{a_{1} a_{1}' b_{1} \\ \sigma_2 \sigma_2'}}  M_{a_1 a_{2}}^{\sigma_2 \ast} W_{b_{1} b_2}^{\sigma_2 \sigma_2'} 
         \left( \sum\limits_{\sigma_1 \sigma_1'}  M_{1 a_1}^{\sigma_1 \ast} W_{1 b_1}^{\sigma_1 \sigma_1'} \Bar{M}_{1 a_1'}^{\sigma_1'} \right) 
       \Bar{M}_{a_1' a_2'}^{\sigma_2'} 
     \right) \cdots \right) \Bar{M}_{a_{L-1}' 1}^{\sigma_L'} \, .
  \end{split}
\end{equation}

The expectation value can then be evaluated by first calculating the expression in the innermost parentheses, which corresponds to a contraction of the tensors associated with the first site.
The parentheses can then be evaluated sequentially until the outermost part has been reached. This procedure can be converted into a set of recursive equations,

\begin{eqnarray}
  \mathbb{L}_{a_0 a_0'}^{b_0} &=& 1 \, , 
    \label{eq:LeftBoundarySite1} \\
  \mathbb{L}_{a_1 a_1'}^{b_1} &=&  \sum\limits_{\sigma_1 \sigma_1'}  M_{1 a_1}^{\sigma_1 \ast} W_{1 b_1}^{\sigma_1 \sigma_1'} \Bar{M}_{1 a_1'}^{\sigma_1'} \, , 
    \label{eq:LeftBoundarySiteI} \\
    & \vdots \nonumber\\
  \mathbb{L}_{a_l a_l'}^{b_l} &=& \sum_{\mathclap{\substack{a_{l-1} a_{l-1}' b_{l-1} \\ \sigma_l \sigma_l'}}}  M_{a_{l-1} a_l}^{\sigma_l \ast} W_{b_{l-1} b_l}^{\sigma_l \sigma_l'} \mathbb{L}_{a_{l-1} a_{l-1}'}^{b_{l-1}} \Bar{M}_{a_{l-1}' a_l'}^{\sigma_l'} \, .
    \label{eq:LeftBoundarySiteL}
\end{eqnarray}
At the last site we eventually obtain

\begin{equation}
  \mathbb{L}_{a_L a_L'}^{b_L} = \langle \Psi \vert \hat{\mathcal{W}} \vert \Bar{\Psi} \rangle \, .
  \label{eq:LastSite}
\end{equation}

The tensor $\mathbb{L}_{a_l a_l'}^{b_l}$ is referred to as the left boundary at site $l$.\cite{Schollwoeck2011_Review-DMRG,keller15,Freitag2020_BookChapter}
The right boundary $\mathbb{R}_{a_{l-1}' a_{l-1}}^{b_{l-1}}$ can be defined analogously as

\begin{equation}
  \mathbb{R}_{a_{l-1}' a_{l-1}}^{b_{l-1}} = \sum_{\mathclap{\substack{a_{l} a_{l}' b_{l} \\ \sigma_l \sigma_l'}}} \Bar{M}_{a_{l-1}' a_l'}^{\sigma_l'}  W_{b_{l-1} b_l}^{\sigma_l \sigma_l'} \mathbb{R}_{a_{l}' a_{l}}^{b_{l}} M_{a_{l-1} a_l }^{\sigma_l \ast} \, .
  \label{eq:}
\end{equation}

Note that a naive evaluation of Eq.~(\ref{eq:LeftBoundarySiteI}) scales as $\mathcal{O}(m^4 \cdot b_l^2 \cdot N^2)$ for site $l$, and therefore the cost of each contraction step is independent of $L$. Accordingly, the overall scaling with $L$ is linear, and not exponential as for full CI.\\

With the definition of the left and right boundaries, the expectation value $\langle \Psi \vert \hat{\mathcal{W}} \vert \Bar{\Psi} \rangle$ can be calculated at any site on the lattice according to

\begin{equation}
  \langle \Psi \vert \hat{\mathcal{W}} \vert \Bar{\Psi} \rangle = \sum\limits_{a_l a_l' b_l} \mathbb{L}_{a_l a_l'}^{b_l} \mathbb{R}_{a_l' a_l}^{b_l} \, .
  \label{eq:FinalExpectationValue}
\end{equation}

It should be noted that the left and right boundaries will not change if the tensor network is modified further to the right or to the left of site $l$, respectively.
Therefore, the boundaries can be stored for a given site, and are only updated once an MPS tensor, which is contained within that boundary, is changed.
This property allows for an efficient sweep-based optimization, as described in the next Section.\\

\subsection{MPS Optimization}
\label{subsec:opt}

\begin{figure}[ht]
	\centering
     \begin{subfigure}[t]{0.4\textwidth}
         \centering
         \includegraphics[height=2.0cm]{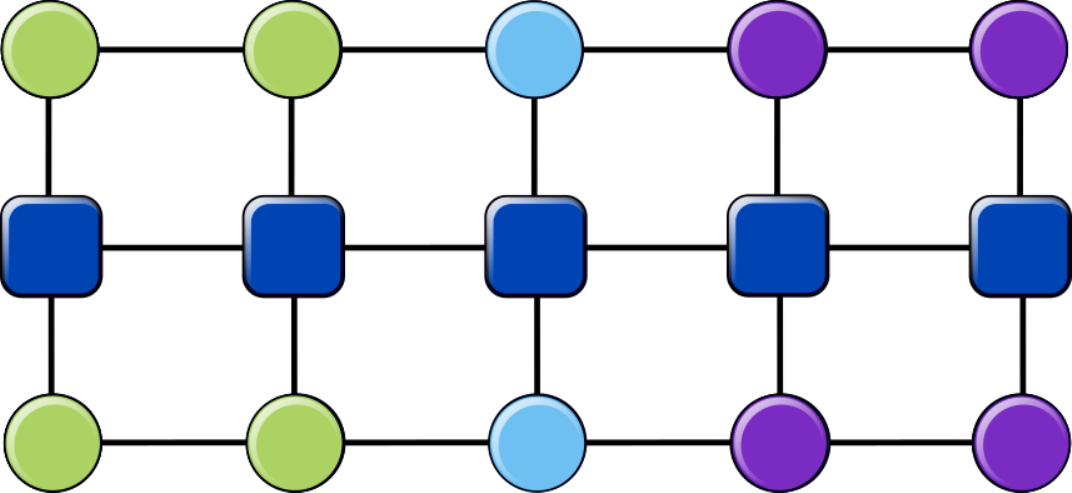}
         \caption{Expectation value without boundaries}
         \label{fig:mps_mpo_mixed}
     \end{subfigure}
     \hspace{1cm}
     \centering
     \begin{subfigure}[t]{0.4\textwidth}
         \centering
         \includegraphics[height=2.0cm]{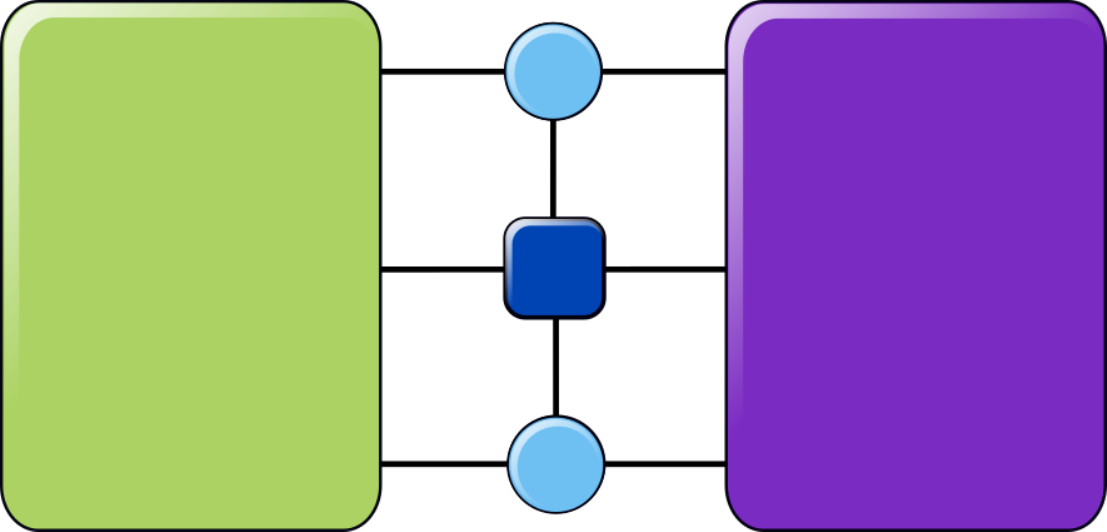}
         \caption{Expectation value with boundaries}
         \label{fig:boundaries}
     \end{subfigure}
     \hfill
     \caption{Tensor network diagrams of the expectation value $\langle \Psi \vert \hat{H} \vert \Psi \rangle$:
     (a) displays the full tensor network contraction according to Eq.~(\ref{eq:expectation_no_boundaries}), while in (b) left and right boundaries are shown as light green and purple boxes containing the left and right contracted portions of the network according to Eq.~(\ref{eq:expectation_boundaries}), respectively.}
\end{figure}

In a variational DMRG calculation, the expectation value $\langle \Psi \vert \hat{\mathcal{H}} \vert \Psi \rangle$ is optimized with respect to the tensor entries of site $l$.
The expectation value can be evaluated at site $l$ based on the left and right boundaries formed from left- and right-normalized matrices (see Fig.~\ref{fig:boundaries}) as

\begin{equation}
  \langle \Psi \vert \hat{\mathcal{H}} \vert \Psi \rangle = \sum\limits_{a_l a_l' b_l} \hspace{20pt} \sum_{\mathclap{\substack{a_{l-1} a_{l-1}' b_{l-1} \\ \sigma_l \sigma_l'}}} 
    M_{a_{l-1} a_l}^{\sigma_l \ast} W_{b_{l-1} b_l}^{\sigma_l \sigma_l'} \mathbb{L}_{a_{l-1} a_{l-1}'}^{b_{l-1}} M_{a_{l-1}' a_l'}^{\sigma_l'}  \mathbb{R}_{a_l' a_l}^{b_l} \, .
  \label{eq:expectation_boundaries}
\end{equation}

To find the optimal MPS parameters for a given site $l$, the gradient of the Lagrangian in Eq.~(\ref{eq:lagrangian}) calculated with respect to the tensor coefficient $M_{a_{l-1} a_l}^{\sigma_l \ast}$ is set to zero. 
This results in the eigenvalue equation

\begin{equation}
  \sum_{\mathclap{\substack{a_{l-1}' a_l' b_{l-1} b_l \\ \sigma_l'}}} W_{b_{l-1} b_l}^{\sigma_l \sigma_l'} \mathbb{L}_{a_{l-1} a_{l-1}'}^{b_{l-1}} M_{a_{l-1}' a_l'}^{\sigma_l'}  \mathbb{R}_{a_l' a_l}^{b_l} = \lambda M_{a_{l-1} a_l}^{\sigma_l}  \, ,
  \label{eq:ev}
\end{equation}
which can be solved with an iterative eigenvalue solver such as the Jacobi--Davidson algorithm.\cite{Davidson1975,sleijpen00}\\
The solution to Eq.~(\ref{eq:ev}) for every coefficient $M_{a_{l-1} a_l}^{\sigma_l}$ returns the optimal MPS tensor $M^{\sigma_l}$ for a given site $l$.
Hence, to fully optimize an MPS, the equation must be solved for all sites of the DMRG lattice in a sequential manner.
The optimization is then repeated iteratively until self-consistency between all tensors is reached. \\

DMRG optimizes the MPS sequentially with a sweeping procedure, as moving from one site to the next enables an optimal recycling of the boundaries.
First, a guess MPS is constructed and brought into the right-canonical form, and the right boundary $\mathbb{R}$ is calculated for every site $l$ on the lattice.
Then, a so-called forward sweep is conducted, where for every site from $l=1$ to $l=L-1$

\begin{enumerate}
  \item the eigenvalue problem of Eq.~(\ref{eq:ev}) is solved for each $M_{a_{l-1} a_l}^{\sigma_l}$ of $\textbf{M}^{\sigma_l}$,
  \item the obtained $\textbf{M}^{\sigma_l}$ tensor is left-normalized (see Eq.~(\ref{eq:LeftNormalization})) to obtain $\textbf{A}^{\sigma_l}$ from $\textbf{U}$,
  \item the MPS coefficients for the next site are updated by multiplying $\textbf{SV}^{\dagger}$ into $\textbf{M}^{\sigma_{l+1}}$,
  \item the left boundary $\mathbb{L}_{a_l a_l'}^{b_l}$ is updated, and the next site to the right on the lattice is selected.
\end{enumerate}

Once the whole lattice has been traversed from the left, the procedure is reversed at site $l=L$.
A backward sweep is conducted, where for every site from $l=L$ to $l=2$

\begin{enumerate}
    \item the eigenvalue problem of Eq.~(\ref{eq:ev}) is solved for each $M_{a_{l-1} a_l}^{\sigma_l}$ of $\textbf{M}^{\sigma_l}$,
    \item the obtained $\textbf{M}^{\sigma_l}$ is right-normalized by performing SVD to obtain $\textbf{B}^{\sigma_l}$ from $\textbf{V}^{\dagger}$,
    \item the MPS coefficients for the neighboring site is updated by multiplying $\textbf{US}$ into $\textbf{M}^{\sigma_{l-1}}$,
    \item the right boundary $\mathbb{R}_{a_l' a_l}^{b_l}$ is updated, and the next site to the left on the lattice is selected.
\end{enumerate}

After the entire lattice has been traversed forth and back, the MPS parameters are optimized further by performing subsequent sweeps until convergence is reached.

\begin{figure}[ht]
  \centering
  \includegraphics[height=2.0cm]{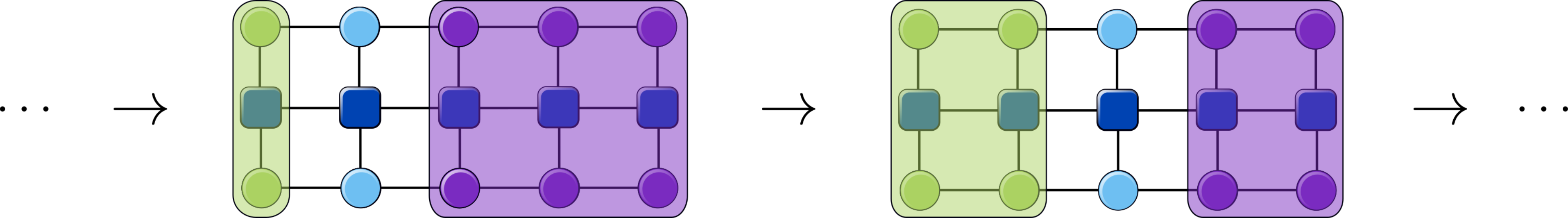}
  \hfill
  \caption{Graphical illustration of two steps in the forward sweep. The shaded boxes represent the left and right boundaries containing the contracted portions of the network in green and purple, respectively. 
  After solving the eigenvalue problem for site $l$, $\textbf{M}^{\sigma_l}$ is left-normalized (hence, turning green), which entails that the tensor for site $l+1$ is also updated and no longer right-normalized (hence, turning blue). 
  The left boundary on site $i$ is updated and the optimization progresses to site $l$+1.}
\end{figure}

The sweeping procedure described above optimizes the tensor entries for a single site at a time.
Alternatively, two sites can be optimized simultaneously with the so-called two-site DMRG variant by introducing two-site tensors for the MPS and for the MPO.\cite{Schollwoeck2011_Review-DMRG}
In this case, the two-site tensor

\begin{equation}
  T_{a_{i-1},a_{i+1}}^{\sigma_i, \sigma_{i+1}} = \sum_{a_i} M_{a_{i-1}, a_i}^{\sigma_i} M_{a_i, a_{i+1}}^{\sigma_{i+1}} \, ,
  \label{eq:TwoSite}
\end{equation}
is constructed and optimized at each microiteration.
The optimized MPS tensors are then obtained from the SVD of the optimized $\textbf{T}^{\sigma_i, \sigma_{i+1}}$ tensor, reshaped as $T_{a_{i-1}\sigma_i, a_i \sigma_{i+1}}$.
The two-site algorithm improves the DMRG convergence and reduces the risk of getting stuck in local minima.
Additionally, it allows for an adaptive setting of the bond dimension $m$, meaning that $m$ can be changed on every microiteration step.
In fact, the number of non-zero singular values of $\textbf{T}^{\sigma_i, \sigma_{i+1}}$ may change during the optimization, and, consequently, so does the bond dimension required for a reliable MPS approximation of the wave function.
While being more stable, two-site DMRG is computationally less efficient than the single-site algorithm as it optimizes larger tensors at each step of the sweeping procedure.
For vibrational applications, the single-site algorithm usually achieves satisfactory convergence and the two-site optimization is applied only to cases where convergence issues are observed.

\subsection{Site Selection and Ordering with Quantum Information Measures}

By truncating the MPS bond dimension $m$, DMRG tames the exponential scaling of the FCI problem, but, at the same time, it limits the extent to which correlation can be encoded in the wave function.
This tradeoff can be counteracted by an optimal arrangement of the sites on the DMRG lattice.
Ideally, a tensor network is constructed in such a way that it directly encodes the correlation structure of the system which it represents.
For an MPS, this translates into placing strongly correlated sites in close vicinity to each other.
Due to the sequential optimization of the MPS with the sweeping procedure, the convergence of DMRG is greatly affected by the ordering of the sites.\\

For an optimal ordering of the sites on the DMRG lattice, the correlation structure of the system at hand has to be determined.
Fortunately, DMRG enables the characterization of the wave function structure with measures from quantum information theory.
These measures can quantify correlation effects in MPSs and can be utilized to optimize the MPS construction.
For this purpose, the system is partitioned into the site of interest $i$ and the environment, in which all other sites are collected. Then, a one-site reduced density matrix (RDM) can be constructed, and a von-Neumann-type entropy can be defined with\cite{Rissler2006}

\begin{equation}
  s(1)_i = - \sum_{\alpha = 1}^{N_i} \omega_{\alpha,i} \ln \omega_{\alpha,i} \, ,
  \label{eq:s1_definition}
\end{equation}
which is referred to as the single-site entropy of the site $i$.
The sum runs over all eigenvalues $\omega_{\alpha,i}$ of the one-site RDM, with $N_i$ being the size of the local basis of site $i$.
Hence, the single-site entropy quantifies how much site $i$ contributes to the deviation of the many-body wave function from a pure state (i.e., one that populates only one of the possible occupation options of a site at all sites).
At the same time, the single-site entropy also provides a measure for the entanglement of site $i$ with all other sites, therefore characterizing the multi-configurational character of site $i$. \\

In an analogous manner, a two-site entropy $s(2)_{ij}$ can be defined for a pair of sites $i$ and $j$.
In this case, the system is defined by orbitals $i$ and $j$, and a two-site reduced density matrix is constructed.
The two-site entropy is then calculated as

\begin{equation}
  s(2)_{ij} = - \sum_{\alpha = 1}^{N_i N_j} \omega_{\alpha,ij} \ln \omega_{\alpha,ij} \, ,
  \label{eq:s2_definition}
\end{equation}
where the sum runs over all $N_i N_j$ eigenvalues $\omega_{\alpha,ij}$ of the two-site RDM, where $N_i$ and $N_j$ are the sizes of the local bases for sites $i$ and $j$, respectively. 
The two-orbital entropy describes how much the combined state of sites $i$ and $j$ is affected by the environment. \\

We define the two-site entropy reduced by the single-site entropies as the mutual information $I_{ij}$ between two sites $i$ and $j$,

\begin{equation}
  I_{ij} = \frac{1}{2} \left( s(1)_i  + s(1)_j -  s(2)_{ij} \right) \left( 1 - \delta_{ij} \right) \, ,
  \label{eq:MutualInformaton}
\end{equation}
which is a measure for how strongly the states of sites $i$ and $j$ are mutually entangled. \\

Based on the von-Neumann-type entropies and the mutual information, the ordering of the sites on the lattice can be optimized and the sites to be included in the MPS can be selected.\cite{Legeza2003_OptimalOrdering}
As with other quantum chemical methods, states which are not strongly correlated do not need to be explicitly included in the wave function expansion.
Hence, both site selection and ordering on the DMRG lattice are greatly facilitated by these measures from quantum information theory.
In practice, these measures can be obtained from a partially converged DMRG calculation with a small maximum bond dimension $m$, because the one- and two-site entropies converge much faster than the energy.
The results from such a calculation are then leveraged for optimizing the MPS site selection and ordering for a subsequent accurate and fully converged DMRG calculation with a sufficiently large bond dimension.

\section{DMRG for Vibrational Problems}

As tensor based methods are powerful tools for a variety of quantum chemical problems, we introduced tensor networks and the DMRG optimization algorithm in a general way in the previous Sections.
In this Section, we will now focus on the vibrational DMRG (vDMRG) theory.\cite{baiardi17,baiardi19c}

\subsection{Vibrational Hamiltonians}

Within the Born--Oppenheimer approximation, molecular vibrations are determined by the potential energy surface (PES) $\mathcal{V}(R)$, which is obtained from the solution of the electronic Schr\"{o}dinger equation at different nuclear configurations $R$.
The PES is, in general, not known exactly and is usually approximated either with a Taylor series expansion around a reference geometry, or with a so-called $n$-mode expansion.\cite{Carter1997_VSCF-CO-Adsorbed,Bowman2003_Multimode-Code,Kongsted2006_nMode,Vendrell2007_ProtonatedWater-15D}
For a vDMRG calculation, the Hamiltonian and, therefore, the PES must be expressed in second-quantized form obtained by projection onto a finite modal basis set.
Depending on the choice for the PES approximation and for the underlying local modal basis, different forms of the second-quantized vibrational Hamiltonian are obtained.
Two quantization choices, namely the canonical quantization\cite{Hirata2014_NormalOrdered} and the $n$-mode representation\cite{christiansen04}, are discussed in the context of the vDMRG method in this Section. \\

\subsubsection{Canonical Quantization}

The canonical quantization is a convenient framework to encode a PES that is expressed as a Taylor series around a reference geometry in terms of the Cartesian normal modes $Q_i$.
The Taylor series expansion of the PES around a stationary point $\textbf{Q}_e$ is given, in atomic units, by

\begin{eqnarray}
  \mathcal{V} &=& \frac{1}{2} \sum_{i=1}^L \left( \frac{\partial^2 V}{\partial Q_i^2} \right)_{\textbf{Q}_e} Q_i^2 
    + \frac{1}{6} \sum_{ijk} \left( \frac{\partial^3 V}{\partial Q_i \partial Q_j \partial Q_k} \right)_{\textbf{Q}_e} 
                  Q_i Q_j Q_k + \; \dots \\
              &=& \frac{1}{2} \sum_{i=1}^L \omega_i^2 Q_i^2 + \frac{1}{6} \sum_{ijk} k_{ijk} Q_i Q_j Q_k+ \; \dots \, , \label{eq:taylor}
\end{eqnarray}
where we omitted the energy at the reference point and note that all first derivatives vanish by definition of $\textbf{Q}_e$.
The harmonic potential is obtained by retaining only the second-order term of Eq.~(\ref{eq:taylor}).
As the Schr\"odinger equation can be solved exactly for the harmonic oscillator Hamiltonian, the corresponding eigenfunctions can be taken as a reference modal basis to expand the PES.
The second-quantization representation of the resulting Hamiltonian can then be obtained by expressing the position operators $Q_i$ in terms of canonical quantization creation and annihilation operators as

\begin{equation}
  Q_i = \frac{1}{\sqrt{2}} \left( \hat{b}_i^+ + \hat{b}_i \right) \, .
  \label{eq:CanonicalQuantization}
\end{equation}
Hence, a pair of second-quantization operators $\hat{b}_i^+$ and $\hat{b}_i$ is introduced for each mode $i$,
whose definition implies that the vibrational wave function is expanded in terms of the harmonic oscillator eigenfunctions.
The canonical quantization also determines the mapping of the Hamiltonian to the vDMRG lattice, namely that each site corresponds to a vibrational normal mode.
As harmonic modes are bosons, the canonical operators fulfill bosonic commutation relations and their application to a given occupation number vector results in:

\begin{eqnarray}
  \hat{b}_i^+ \vert n_1, ... \, , n_i, ... \, , n_L\rangle 
    &=& \sqrt{n_i + 1} \; \vert n_1, ... \, , n_i + 1, ... \, , n_L\rangle \, ,
    \label{eq:CreationOperator} \\
  \hat{b}_i \vert n_1, ... \, , n_i, ... \, , n_L\rangle 
    &=& \sqrt{n_i} \; \vert n_1, ... \, , n_i - 1, ... \, , n_L\rangle \, .
    \label{eq:AnnihilationOperator}
\end{eqnarray}
The dimension of the local basis and, hence, the maximum occupation of each site is, in principle, unbounded.
In practice, however, a threshold $N_i$ is set to limit the size of the local basis at site $i$, as highly excited modes usually have a negligible effect on the anharmonic vibrational wave function.

\begin{figure}[htbp!]
  \centerline{\includegraphics[width=0.5\textwidth]{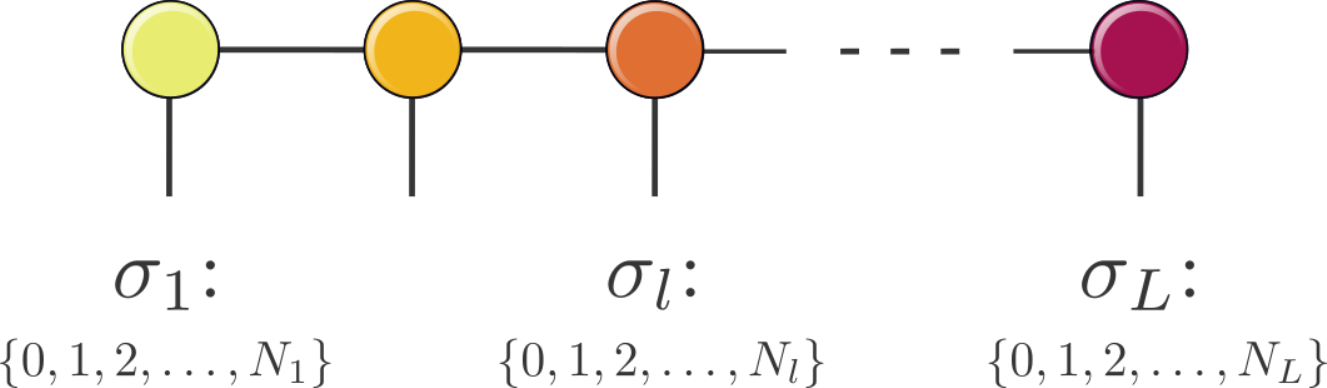}}
  \caption{Vibrational MPS in the canonical quantization basis. Each site corresponds to a certain vibrational mode, as highlighted by the different colors. For each vibrational mode $i$, the local basis consists of all vibrational excitations of that mode up to the defined upper bound $N_i$.}
  \label{fig:canonical}
\end{figure}

While the canonical quantization provides an efficient description of weakly anharmonic systems, it suffers from two severe limitations in the case of strong anharmonicity.
First, the Taylor series expansion of the PES does not accurately parametrize potentials without a clear single reference geometry, such as double-wells.
Second, the harmonic oscillator eigenfunctions as a basis set do not represent strongly anharmonic wave functions well.
These drawbacks can be overcome with the $n$-mode quantization, which allows for a more flexible representation of the vibrational Hamiltonian.

\subsubsection{$n$-mode Quantization}

In the $n$-mode picture, the PES is approximated with a many-body expansion in which terms are grouped together based on the number of modes that are coupled together.\cite{Carter1997_VSCF-CO-Adsorbed,Bowman2003_Multimode-Code,Kongsted2006_nMode,Vendrell2007_ProtonatedWater-15D}
The potential is written as

\begin{equation}
  \mathcal{V} = \sum\limits_{i=1}^M \mathcal{V}^{[i]}(Q_i) + \sum\limits_{i<j} \mathcal{V}^{[ij]}(Q_i, Q_j) 
              + \sum\limits_{i<j<k} \mathcal{V}^{[ijk]}(Q_i, Q_j, Q_k) + \ldots
  \label{eq:nMode}
\end{equation}
where $M$ is the number of modes.
The one-body term $\mathcal{V}^{[i]}(Q_i)$ contains the variation of the potential upon change of the $i$-th coordinate $Q_i$.
Analogously, the two-body term $\mathcal{V}^{[ij]}(Q_i, Q_j)$ represents the variation of the PES for the simultaneous change of two coordinates, $Q_i$ and $Q_j$ and, therefore, increases with the degree of coupling between modes $i$ and $j$.
In contrast to the electronic case, interactions are not limited to one- and two-body terms, as higher-order coupling terms may be relevant.
In practice, however, it is often sufficient to only include low-order terms as they contain most of the correlation between the modes. \\

The $n$-mode expansion of the PES supports a second-quantization form that is based on generic basis sets.
This allows one to choose a basis set that is optimized to yield compact CI wave functions, such as modals obtained from a VSCF calculation.\cite{bowman86,hansen10}
In the $n$-mode picture, a pair of creation and annihilation operators must then be introduced for each of the modal basis functions $k_i$ of each mode $i$, as opposed to the canonical quantization, where only a single pair of operators is required per mode.
Hence, an occupation number vector is given as $| \textbf{n} \rangle = | n_1^1 , \, ... \, , n_{N_1}^1, \, ... \, , n_1^i, \, ... \, , n_{N_i}^i, \, ... \, , n_1^M , \, ... \, , n_{N_M}^M \rangle$, where $n_{k_i}^i$ represents the occupation of the $k_i$-th basis function associated with the $i$-th mode.
The corresponding creation $\hat{a}_{k_i}^+$ and annihilation $\hat{a}_{k_i}$ operators are defined as

\begin{eqnarray}
  \hat{a}_{k_i}^+ | \textbf{n} \rangle
    &=&  | \textbf{n} + \boldsymbol{1}^i_{k_i} \rangle  \, \delta_{0,n_{k_i}^i} \, ,
    \label{eq:CreationOperatorNMode} \\
  \hat{a}_{k_i} | \textbf{n} \rangle 
    &=&  | \textbf{n} - \boldsymbol{1}^i_{k_i} \rangle  \, \delta_{1,n_{k_i}^i} \, ,
    \label{eq:AnnihilationOperatorNMode}
\end{eqnarray}

where $|\boldsymbol{1}^i_{k_i} \rangle$ is the ONV with zero entries except $n_{k_i}^i=1$.

The second-quantized $n$-mode potential results in\cite{christiansen04}
\begin{equation}
  \mathcal{V} = 
      \sum\limits_i \sum\limits_{k_i,h_i} \langle k_i \vert \mathcal{V}_i (Q_i) \vert h_i \rangle a_{k_i}^+ a_{h_i} 
    + \sum\limits_{i,j} \sum\limits_{k_i,h_i} \sum\limits_{k_j,h_j} \langle k_i k_j \vert \mathcal{V}_{ij} (Q_i, Q_j) 
      \vert h_i  h_j\rangle a_{k_i}^+ a_{k_j}^+ a_{h_j} a_{h_i} \, .
  \label{eq:SQPotentialNMode}
\end{equation}
As each mode is described by a single basis state in each many-body basis function, exactly one of the corresponding modals is occupied in an ONV, whereas all others are unoccupied.
Each individual site can only be in one of two states: either it is occupied, if the mode is represented by that specific modal, or it is unoccupied, if the mode is in another basis state.
Therefore, the creation and annihilation operators $a_{k_i}^+$ and $a_{h_i}$ always occur in pairs for a given mode $i$ in Eq.~(\ref{eq:SQPotentialNMode}), as mode $i$ is excited from modal $h_i$ to modal $k_i$.\\

\begin{figure}[htbp!]
  \centerline{\includegraphics[width=0.8\textwidth]{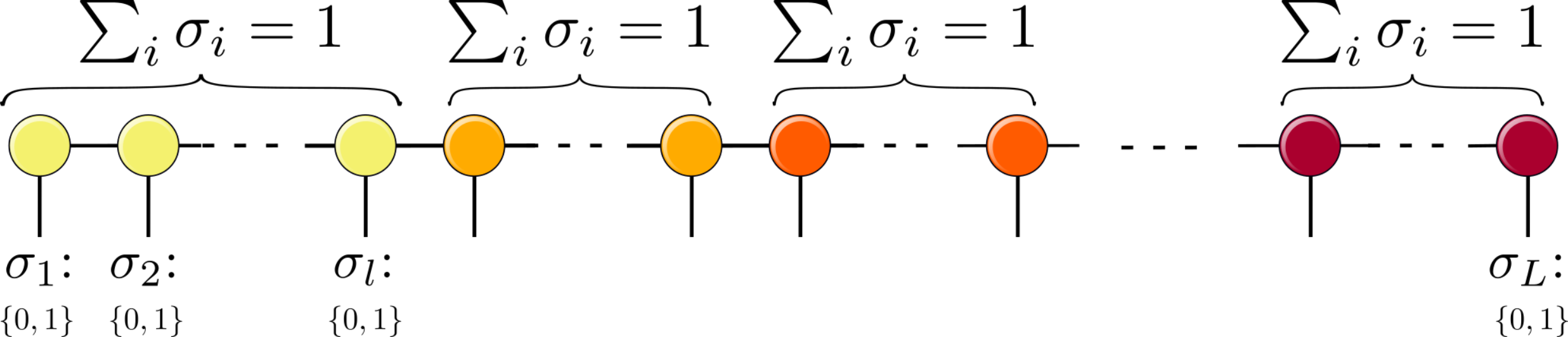}}
  \caption{Vibrational MPS in the $n$-mode quantization. Each vibrational mode is mapped to a number of sites, displayed here as circles of the same color.
  Each site corresponds to one of the modals of that specific mode.
  Each mode is described by only one modal, so that exactly one of their modals has to be in an occupied state, with all others being unoccupied.
  Hence, the local basis of each site is either unoccupied ($\sigma_i = 0$) or occupied ($\sigma_i = 1$), with the sum over all sites belonging to the same mode being $1$.}
  \label{fig:nmode}
\end{figure}

For a vDMRG calculation in the $n$-mode picture, a vibrational mode is no longer mapped to a single site on the lattice, but to a number $N_i$ of sites, one for each local basis function.
Therefore, in the $n$-mode quantization the lattice size $L$ no longer corresponds to the number of modes, but to the total number of modals of all modes.
The $n$-mode lattice is graphically represented in Fig.~\ref{fig:nmode}.
For a given vibrational problem with $M$ modes, each with $N_i$ basis functions, the canonical quantization results in a significantly smaller vDMRG lattice with size $L$=$M$, whereas the $n$-mode lattice has size $L = \sum_i N_i$.
The increase in computational cost with the lattice size $L$ is counterbalanced by the smaller local basis in the $n$-mode picture, which decreases the cost of a single mictroiteration of the vDMRG algorithm.
In general, the $n$-mode quantization is advantageous for strong anharmonicity, because it allows for a more flexible description of the vibrational states and is able to accurately parametrize strongly coupled and highly anharmonic modes.

\subsection{Initial Guess and Bond Dimension}

To initiate a vDMRG calculation, a guess MPS has to be chosen as a starting point for the optimization.
While the DMRG algorithm may converge independently of the initial guess MPS, the convergence rate to the global minimum can be enhanced significantly by appropriately choosing the starting guess.
In practice, the MPS is commonly initialized with the mean-field reference state.
For the ground state, this corresponds to all modes being in the lowest-energy basis state in the canonical quantization, while in the $n$-mode picture each mode is initialized in the lowest-energy modal. \\

The most vital parameter of a vDMRG calculation is the bond dimension $m$, as it controls the tradeoff between computational complexity and the degree of approximation of the full CI wave function.
The minimal bond dimension required to accurately represent a many-body wave function with an MPS depends on the extent of its entanglement.
As the correlation within a system is usually not known prior to the calculation, the energy convergence with respect to the bond dimension has to be monitored.
In practice, this is done by performing several vDMRG calculations with different bond dimensions and extrapolating the energy of the optimized MPS.
By virtue of the variational principle, a lower final energy consistently indicates a better approximation to the ground state wave function.
Therefore, a vDMRG calculation will be converged if the energy does not change when $m$ is enlarged.
In contrast to electronic structure DMRG calculations, converged vDMRG results can be obtained already with relatively small bond dimensions, as demonstrated in Section~\ref{sec:ex_gs}.

\subsection{Example - Anharmonic ZPVE of Ethylene}
\label{sec:ex_gs}

vDMRG can be straightforwardly applied to calculate the anharmonic zero point vibrational energy (ZPVE) of molecular systems.
As discussed above, vDMRG convergence must be monitored both with respect to the number of sweeps and to the bond dimension.
To illustrate this, we apply vDMRG to ethylene with the sixth-order Taylor expanded PES reported in Ref.~\citenum{Berkelbach2021_HBCI-Vibrational} that we encode as an MPO based on the canonical quantization picture.
For all examples, we rely on the single-site DMRG optimization algorithm implemented in the \texttt{QCMaquis} software module.\cite{baiardi17,qcmaquis}

\begin{figure}[ht!]
	\centering
    \includegraphics[width=0.8\textwidth]{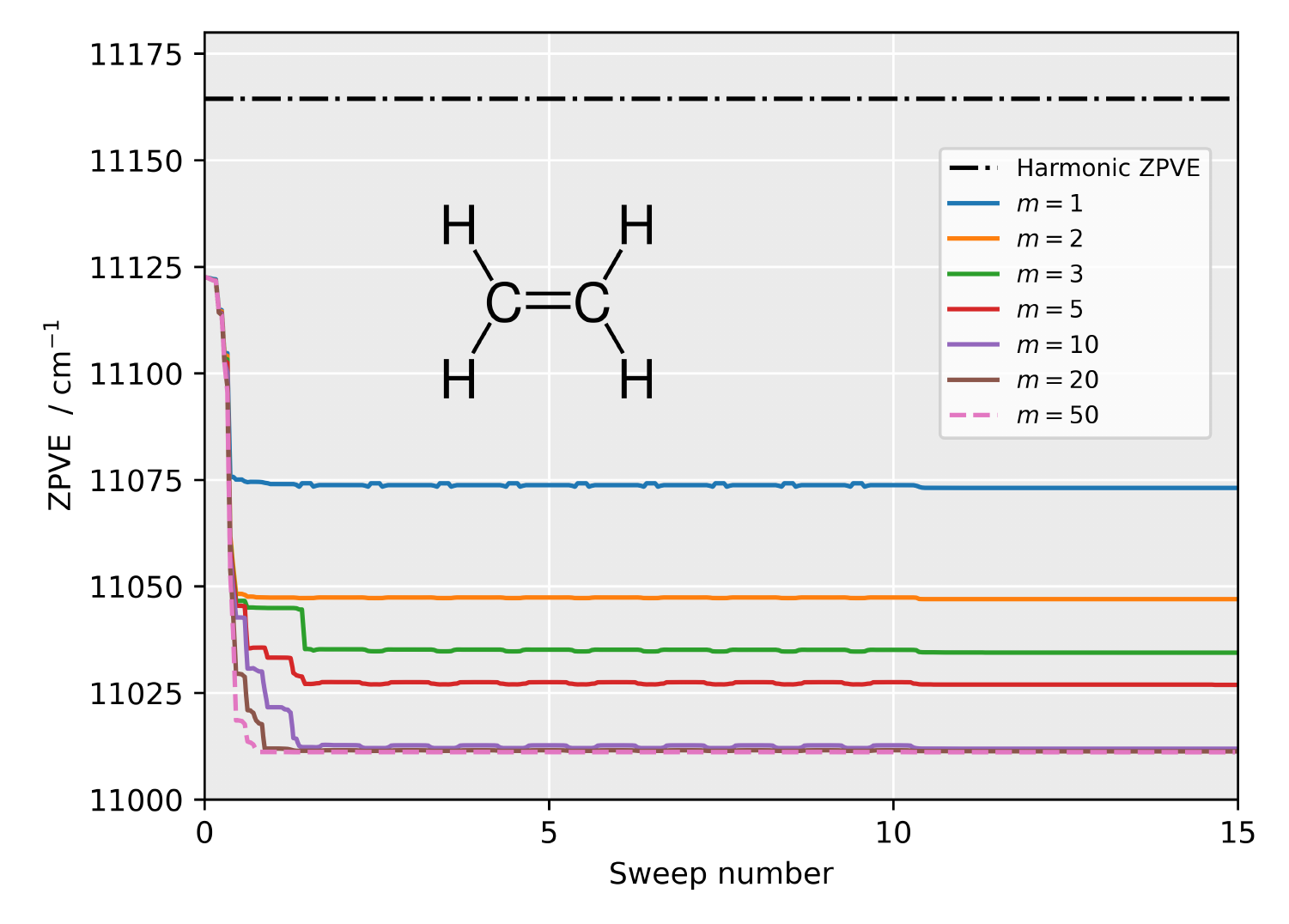}
    \hfill
    \caption{vDMRG energy convergence with respect to the number of sweeps of the ground state of ethylene for a range of bond dimensions $m$.}
    \label{fig:ethylene_gs}
\end{figure}

All calculations reported in Fig.~\ref{fig:ethylene_gs} are initialized in the harmonic ground state.
Their initial energy, the anharmonic ZPVE of the harmonic ground-state wave function, is significantly lower than the harmonic reference ZPVE.
During the vDMRG calculation, the anharmonic ZPVE decreases rapidly during the first few sweeps as the MPS is optimized.
For $m$=1 (blue line in Fig.~\ref{fig:ethylene_gs}), the calculation converges to the anharmonic mean-field wave function.
By increasing the bond dimension $m$, the MPS can account for more correlations between the different modes and the anharmonic ZPVE decreases accordingly.
No significant change in the final energy is observed for bond dimensions higher than $m$=20, indicating convergence of the vDMRG calculations with respect to $m$.
For the first $10$ sweeps, small oscillations are observed in the energy convergence.
This phenomenon is observed because the MPS is perturbed based on the subspace expansion algorithm\cite{Hubig2015_SingleSite} during the first 10 sweeps in order to avoid converging to local minima of the energy functional. \\

As can be seen in Fig.~\ref{fig:ethylene_gs}, the ZPVE of ethylene decreases significantly if anharmonic effects are taken into account.
vDMRG converges rapidly, both with the number of sweeps as well as with the bond dimension $m$, so that the energies are converged to sub-cm$^{-1}$ accuracy for ethylene already after a few sweeps.

\section{Excited State DMRG}

The DMRG algorithm is based on the variational principle and is, therefore, tailored to the optimization of ground-state wave functions.
When applied to vibrational problems, vDMRG returns the anharmonic ZPVE $E_0$.
The quantities of interest in vibrational spectroscopy are, however, not the ZPVEs, but the transition frequencies $\nu_k$, which are calculated from excited state energies $E_k$ as

\begin{equation}
  h \nu_k = E_k - E_0 \, .
  \label{eq::VibrationalEnergy}
\end{equation}
Therefore, excited state vDMRG algorithms are needed to calculate $\nu_k$, of which various have been developed in the past few years.
In this Section, we briefly introduce some of them which target excited state solutions of the time-independent Schr\"{o}dinger equation.
We note that while many of these approaches have originally been developed for vibrational problems, they also apply to electronic structure problems.
In the subsequent Section, we will then review an entirely different approach based on quantum dynamics, which solves the time-dependent Schr\"{o}dinger equation with a vDMRG-based algorithm to calculate spectra in the time domain.

\subsection{Excited-State Targeting by Constrained Optimization with \\ vDMRG[ortho]}

The simplest extension of the regular vDMRG algorithm for vibrational spectroscopy targets excited states based on a constrained optimization.\cite{baiardi17}
As all non-degenerate eigenstates of a Hermitian operator are mutually orthogonal, excited states can still be targeted with a variational calculation provided that the optimization is restricted to the variational space orthogonal to all lower-energy states.
In the vDMRG[ortho] method,\cite{McCulloch2007_FromMPStoDMRG,keller15} excited states are calculated sequentially by optimizing the current state under the constraint that the MPS is orthogonal to the previously calculated MPS of all lower-lying states.\\

\begin{figure}[ht]
  \centering
  \begin{subfigure}[t]{0.4\textwidth}
    \centering
    \includegraphics[height=1.50cm]{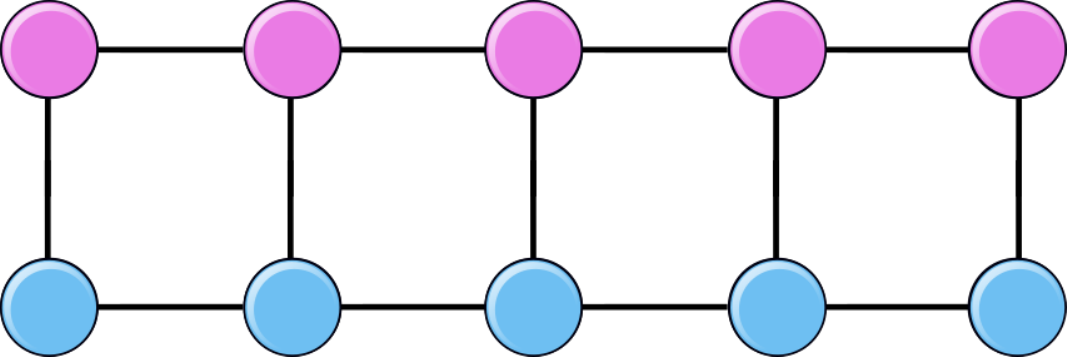}
    \caption{Overlap without boundaries}
    \label{fig:mps_overlap}
  \end{subfigure}
  \hspace{1cm}
  \centering
  \begin{subfigure}[t]{0.4\textwidth}
    \centering
    \includegraphics[height=1.7cm]{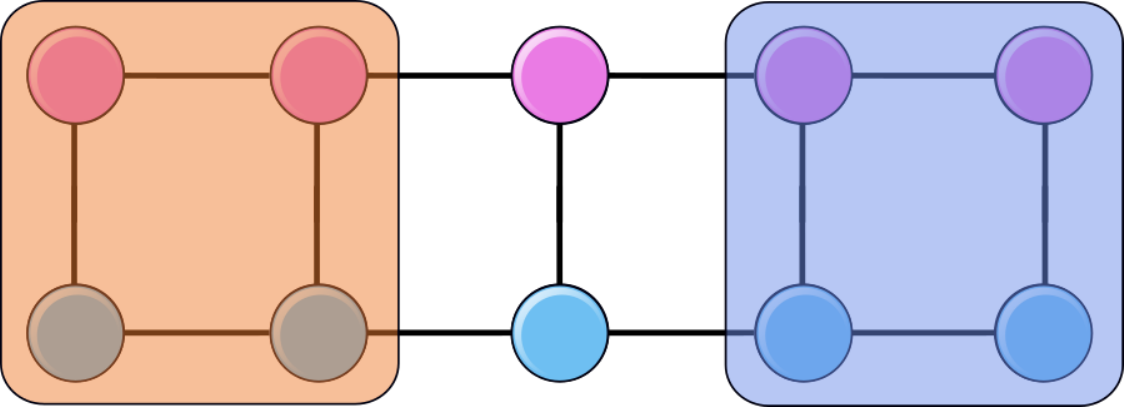}
    \caption{Overlap with boundaries}
    \label{fig:overlap_boundaries}
  \end{subfigure}
  \hfill
  \caption{Tensor network diagrams of the overlap $\langle \Phi \vert \Psi \rangle$.
  (a) The overlap of two different MPS, as highlighted by their different coloring, is evaluated with a naive contraction of the entire network.
  (b) The overlap evaluation can be greatly facilitated by calculating and storing the boundaries, which collect the precontracted portions of the network, as illustrated by the shaded boxes.
  In vDMRG[ortho], the overlap boundaries are stored and updated for every site $l$ on the lattice in analogy to the expectation value boundaries discussed in Section~\ref{subsec:expval}.}
  \label{fig:overlap_2mpos}
\end{figure}

While the vDMRG[ortho] method allows for a straightforward calculation of low-energy excited states, there will be significant drawbacks if it is applied to higher-lying excited states.
As the algorithm requires all lower-lying states as input, not only the calculation time increases, but also the error which is introduced into the calculation.
In fact, the constrained optimization becomes increasingly unstable as the number of constraints grows, especially for small bond dimensions.
Possible remedies for this instability are to increase the bond dimension or to track a specified state during the calculation with an overlap criterion introduced in the next Section.

\subsection{Enhancing Convergence by Root Homing: vDMRG[maxO]}

\begin{figure}[ht]
	\centering
    \includegraphics[height=3.0cm]{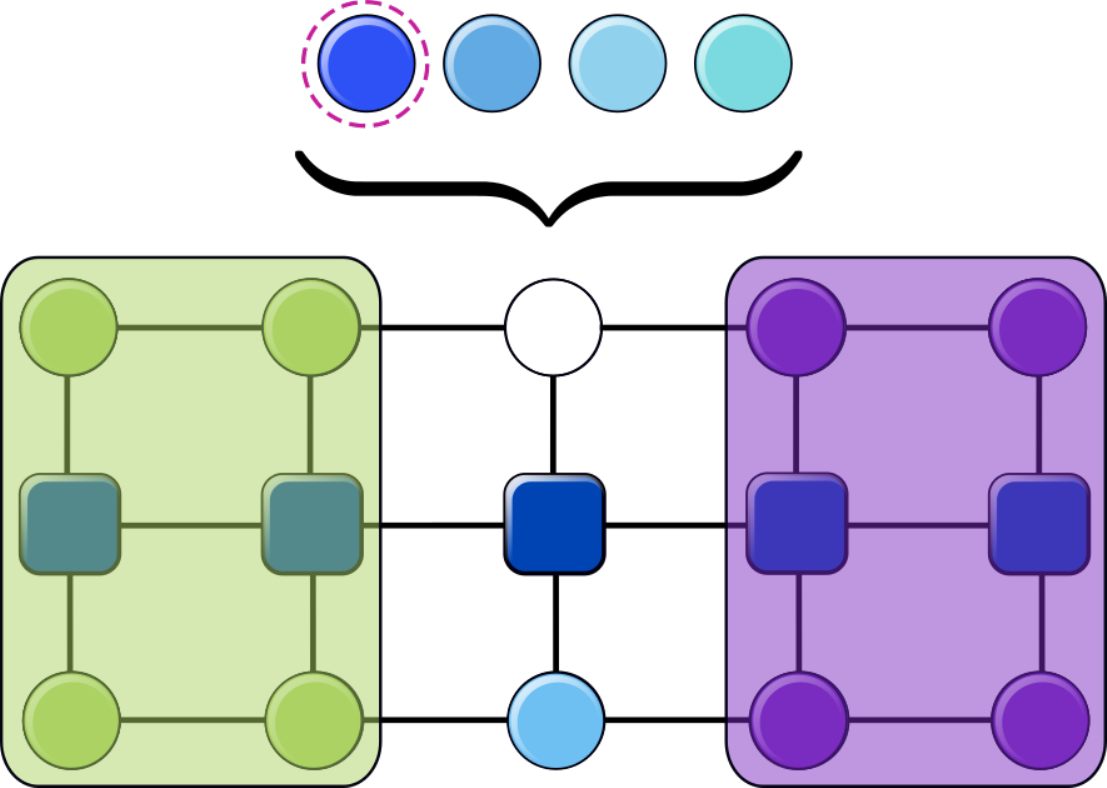}
    \hfill
    \caption{Graphical illustration of the root homing procedure.
    At each microiteration step of the optimization procedure, multiple eigenpairs of Eq.~(\ref{eq:ev}) are obtained, as illustrated by the four circles.
    These are sorted according to their overlap with the reference MPS, as visualized by the darker to lighter blue shading of the tensors.
    In a vDMRG[maxO] calculation, the tensor which results in the largest overlap with the reference MPS, highlighted by the red circle, is selected and propagated during the subsequent MPS optimization.}
    \label{fig:root}
\end{figure}

Another extension of vDMRG resting on the variational energy minimization is the vDMRG[maxO] method.\cite{baiardi19c}
In this variant, the MPS optimization is steered by a maximum overlap criterion.
During the DMRG sweeping procedure, the eigenvalue equation of Eq.~(\ref{eq:ev}) is solved at each microiteration step and yields multiple eigenpairs.
In standard DMRG, the tensor with the lowest energy is selected to propagate the ground-state wave function.
This selection scheme can be adapted to follow a different eigenpair based on the maximum overlap criterion instead of the minimal energy criterion, as illustrated in Fig.~\ref{fig:root}.
This procedure is also known as root homing,\cite{Butscher1976_RootHoming} as a specific root of the eigenvalue equation is selected. 
For a vDMRG[maxO] calculation, a reference MPS has to be chosen with respect to which the overlap is calculated.
The predetermined reference MPS is often selected from the eigenstates of the non-interacting Hamiltonian, corresponding to the excited state mean-field reference.
Alternatively, the reference MPS can be obtained from a preliminary vDMRG calculation with a smaller bond dimension $m$.
To allow for an efficient overlap calculation, partial overlaps are constructed as illustrated in Fig.~\ref{fig:overlap_boundaries}, stored for each site, and updated during the sweeping procedure. \\

The vDMRG[maxO] method can be applied to directly target excited states by choosing a reference MPS which approximates the state of interest.
vDMRG[maxO] can also be combined with other excited state methods to ensure convergence to the desired state even in regions with a high density of states.
Particularly appealing is the combination of the maximum overlap criterion with the vDMRG[S\&I] method which is introduced in the next Section.

\subsection{Auxiliary operator based algorithms: vDMRG[S\&I] and vDMRG[f]}

\begin{figure}[htbp!]
  \centerline{\includegraphics[width=0.4\textwidth]{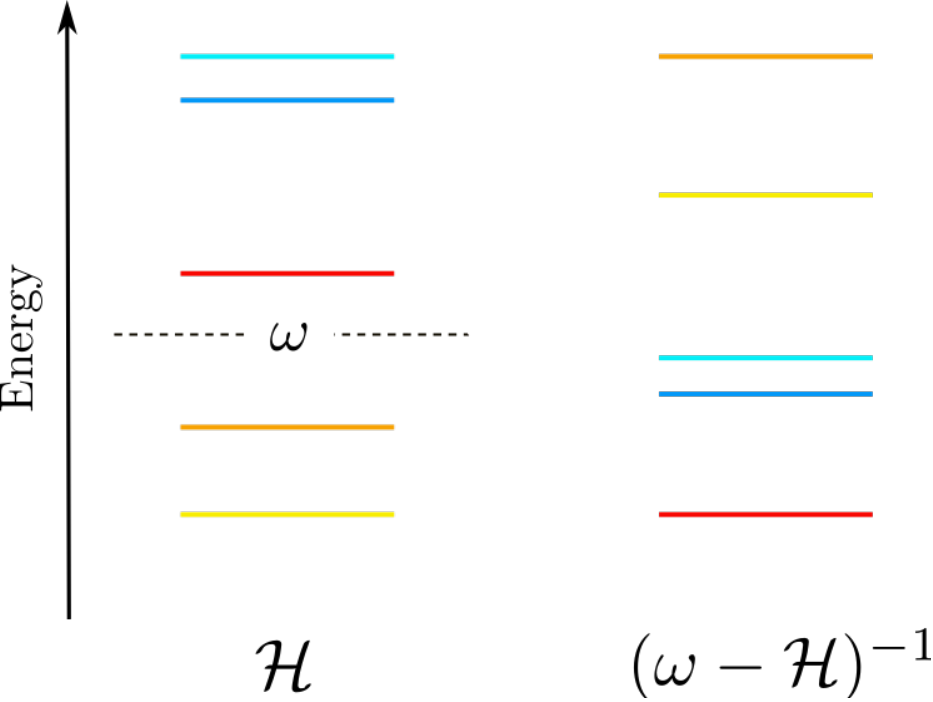}}
  \caption{Illustration of the energy eigenvalues of the shifted and inverted Hamiltonian $\Omega_{\omega}$.
  The red state, which is the first with an energy above the shift $\omega$, is mapped to the ground state of the auxiliary operator $\Omega_{\omega} = \left( \omega - \mathcal{H} \right)^{-1}$ and can, therefore, be targeted by variational optimization with vDMRG[S\&I].}
  \label{fig:sandi}
\end{figure}

As an alternative to constraining the MPS optimization, the vDMRG calculation can be steered by mapping the Hamiltonian to an auxiliary operator to directly optimize an excited state with ground state DMRG.
For instance, the vDMRG[S\&I] algorithm applies conventional vDMRG to the shifted-and-inverted auxiliary Hamiltonian,$\Omega_{\omega}$\cite{baiardi19c}
\begin{equation}
  \Omega_{\omega} = \left( \omega - \mathcal{H} \right)^{-1} \, ,
  \label{eq:SandI}
\end{equation}
where $\omega$ is an energy shift parameter.
The smallest eigenvalue of $\Omega_{\omega}$ corresponds to the first eigenvalue of $\mathcal{H}$ with an energy greater than $\omega$.
Hence, a ground state optimization applied to $\Omega_{\omega}$ yields the excited state directly above $\omega$, as illustrated in Fig.~\ref{fig:sandi}.

The vDMRG[S\&I] approach first requires an energy shift $\omega$ so that the algorithm can then optimize a specific excited state.
Since the energy of the target state is, in general, not known \textit{a priori}, a guess based on experimental values or results from previous calculations can to be made.
This becomes an issue in regions with a high density of states, as the energy shift $\omega$ must be chosen with high accuracy to ensure that vDMRG[S\&I] converges to the desired eigenstate.
Fortunately, the vDMRG[S\&I] method can be made more robust by combining it with the maximum overlap criterion.
However, another major issue remains.
Encoding the auxiliary operator $\Omega_{\omega}$ as an MPO is a challenging numerical problem because, even though $\mathcal{H}$ is represented as a compact MPO, its inverse may be encoded by an arbitrarily complex MPO representation.
Algorithms that approximate the inverse of an MPO have been developed,\cite{Oseledets2012_ALS,Lubich2014_TimeIntegrationTT} but they have never been applied to quantum chemical problems so far.
The MPO inversion can be avoided by applying the S\&I transformation to the local, site-centered operator defined by Eq.~(\ref{eq:ev})\cite{Dorando2007_Targeted,baiardi19b} based on the theory of harmonic Ritz values.\cite{sleijpen00}
However, if the S\&I transformation is applied after the eigenvalue equation is projected onto a given site, the method's variationality and, therefore, its stability, will not be ensured anymore.\cite{Oseledts2016_VDMRG}
For this reason, vDMRG[S\&I] may not converge if applied to optimize high-energy states, even if it is combined with vDMRG[maxO].

\begin{figure}[htbp!]
  \centerline{\includegraphics[width=0.3\textwidth]{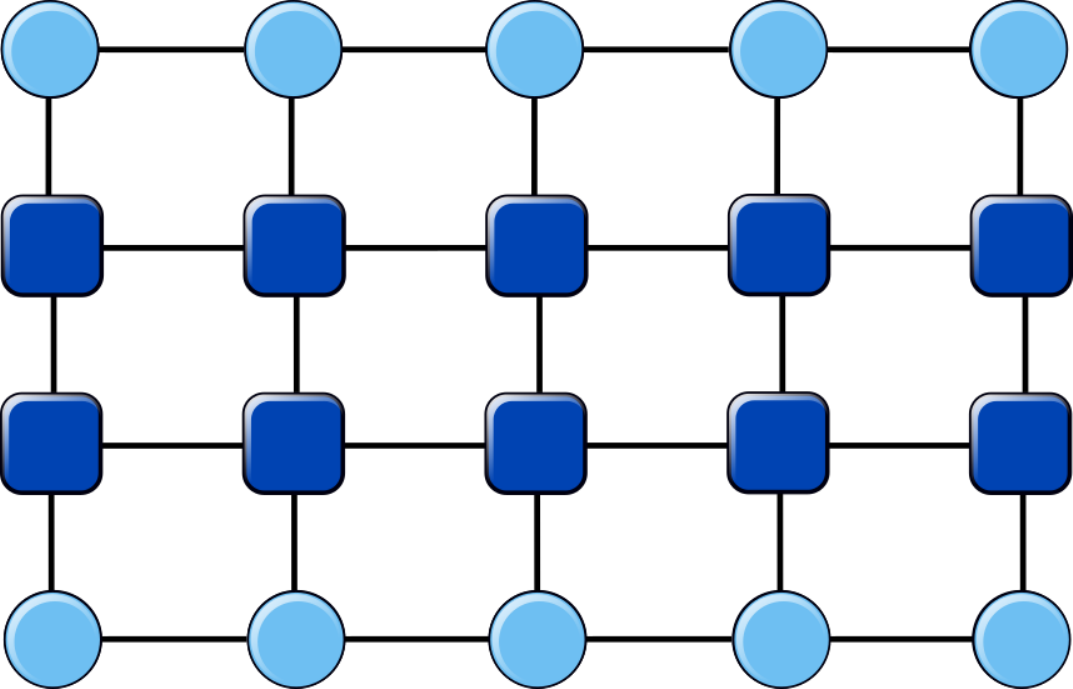}}
  \caption{Tensor network diagram for the expectation value of the folded Hamiltonian. Note that the MPO has to be applied twice onto the MPS to account for the squared operator in Eq~(\ref{eq:Folded}).}
  \label{fig:folded}
\end{figure}

The S\&I transformation is not the only one fulfilling the requirement that the lowest-energy eigenfunction of the corresponding auxiliary operator is the excited state with energy closest to the shift parameter $\omega$.
For instance, the folded operator that is defined as

\begin{equation}
  \Theta_{\omega} = \left( \omega - \mathcal{H} \right)^{2} \, ,
  \label{eq:Folded}
\end{equation}
also satisfies this condition, and is the auxiliary operator employed in folded vDMRG (vDMRG[f]).
On the one hand, vDMRG[f] has the key advantage that the operator defined in Eq.~(\ref{eq:Folded}) can be straightforwardly encoded as an MPO.
On the other hand, evaluating the expectation value in vDMRG[f] is computationally expensive due to the need of contracting the MPS twice with the MPO to account for the square in Eq.~(\ref{eq:Folded}), as illustrated in Fig.~\ref{fig:folded}.
The key advantage is that vDMRG[f] allows for a more stable optimization than vDMRG[S\&I] since it preserves the method's variationality and, therefore, ensures a smooth convergence to the target state for any energy range.

\subsection{Inverse Power Iteration with MPS: the vDMRG[IPI] Algorithm}

An appealing alternative to variational excited-state vDMRG variants are methods relying on projection-based algorithms, such as the vDMRG[IPI] approach\cite{Oseledts2016_VDMRG} that combines DMRG with the inverse power iteration (IPI) method.\cite{saad11}
In IPI, the shift-and-invert operator is repeatedly applied onto a guess wave function,
\begin{equation}
  \vert \Psi_k \rangle = \left( \mathcal{H} - \omega \right)^{-1} \vert \Psi_{k-1} \rangle \, ,
  \label{eq:ipi1}
\end{equation}
where $k$ labels the iteration step in this recursive equation.
Starting from an initial guess $\vert \Psi_0 \rangle$, the wave function $\vert \Psi_k \rangle$ converges with increasing $k$ towards the eigenfunction of the Hamiltonian with the energy closest to the shift $\omega$.
The explicit inversion of the Hamiltonian can be avoided by solving the linear system

\begin{equation}
  \Gamma_{\omega} \vert \Psi_k \rangle = \vert \Psi_{k-1} \rangle \,
  \label{eq:ipi}
\end{equation}
for a normalized wave function $\vert \Psi_k \rangle$, where we define the projection operator as $\Gamma_{\omega} = \mathcal{H} - \omega$.
In vDMRG[IPI], the projection operator $\Gamma_{\omega}$ is encoded as an MPO and the wave function $\vert \Psi_k \rangle$ is approximated as an MPS with a fixed bond dimension $m$.
Unfortunately, the application of an MPO onto an MPS increases the bond dimension of the latter.\cite{Schollwoeck2011_Review-DMRG}
Therefore, Eq.~(\ref{eq:ipi}) can only be solved approximately with DMRG by calculating the optimal MPS of a given bond dimension $m$ with the least squares method.
Therefore, during a vDMRG[IPI] calculation the functional

\begin{eqnarray}
O_{\omega} \left[ \Psi_{k} \right] &=& 
    \left\| \Gamma_{\omega} | \Psi_{k} \rangle - | \Psi_{k-1} \rangle \right\|^2 \\
      &=& \langle \Psi_{k} | \Gamma_{\omega}^2 | \Psi_{k} \rangle 
       - 2 \, \text{Re} \left( \langle \Psi_{k}| \Gamma_{\omega} | \Psi_{k-1} \rangle \right)
       + \langle \Psi_{k-1}| \Psi_{k-1} \rangle \nonumber \\
       &\approx& \langle \Psi_k \vert \Gamma_{\omega} \vert \Psi_k \rangle 
    - 2 \langle \Psi_{k-1} \vert \Gamma_{\omega} \vert \Psi_k \rangle
 \label{eq:residualMinimum}
\end{eqnarray}
is minimized with the alternating least squares algorithm by optimizing one site at a time in a sweep-like fashion. \\

For vDMRG[IPI] calculations, both an energy shift $\omega$ and a starting guess $\vert \Psi_0 \rangle$ must be specified.
While the algorithm is robust with respect to the accuracy of the energy guess $\omega$, the choice for the initial state is crucial to ensure stable convergence to the correct eigenstate.
In practice, robust excited-state targeting can be achieved if, for the calculation of the $n$-th fundamental frequency, the wave function is initialized with the $n$-th mode in the first excited state, while all other modes remain in the vibrational ground state. 

\subsection{Towards large-scale excited-state DMRG: vDMRG[FEAST]}

A common issue of all vDMRG excited state methods that rely on an energy shift to target a specific state is their sensitivity towards the shift parameter $\omega$, especially within highly dense energy ranges.
To accurately choose $\omega$ \textit{a priori} can be a tremendous challenge, and therefore, an automated setting of the energy shift would greatly facilitate excited-state calculations.
A particularly promising method in this respect is the vDMRG[FEAST] approach,\cite{Baiardi2021_DMRG-FEAST} where the FEAST algorithm\cite{polizzi09} is combined with vDMRG.
vDMRG[FEAST] is a robust and efficient method to calculate excited states wave functions encoded as MPS and has three key advantages compared to the previously introduced excited state variants: firstly, the energy shift parameter $\omega$ is determined automatically within the FEAST algorithm and therefore need not be set externally. 
Secondly, linear equations are solved, as in DMRG[IPI], instead of more involved eigenvalue problems.
Finally, the linear systems to be solved in vDMRG[FEAST] are independent of one another and can, therefore, be solved in parallel. \\

The FEAST algorithm is an iterative subspace diagonalization algorithm that can be applied to solve generalized eigenvalue problems.
FEAST relies on a simultaneous optimization of all eigenfunctions with an energy within a given energy target interval by applying a projection operator onto a set of guess wave functions.
Let $I_E = [ E_{\text{min}}, E_{\text{max}} ]$ be an energy interval that contains $M$ eigenvalues of the Hamiltonian $\mathcal{H}$ with eigenfunctions $I_{M} = \{\Psi^{(1)}, \dots,\Psi^{(M)}\}$.
By leveraging the Cauchy integral theorem, the projector $\mathcal{P}_{M}$ onto the space spanned by the $I_M$ functions can be expressed as a complex contour integral with

\begin{equation}
  \mathcal{P}_{M} = \sum_{i=1}^{M}  \vert \Psi^{(i)} \rangle \langle \Psi^{(i)} \vert
                  = \frac{1}{2\pi \mathrm{i}} \oint_{\mathcal{C}} (z - \mathcal{H})^{-1} dz \, .
  \label{eq:FEAST_Projector}
\end{equation}

Starting from a linearly independent set of $M$ guess states $\{\Phi_\text{guess}^{(1)}, \ldots, \Phi_\text{guess}^{(M)}\}$, a basis of the subspace spanned by $I_M$ can be defined as $S_M \{ \mathcal{P}_{M} \Phi_\text{guess}^{(1)}, \ldots, \mathcal{P}_{M} \Phi_\text{guess}^{(M)} \}$.
The eigenfunctions within the interval $I_E$ are then obtained by solving the original eigenvalue problem in the subspace spanned by $S_M$.
Calculating the projector defined in Eq.~(\ref{eq:FEAST_Projector}) requires both inverting the Hamiltonian and evaluating exactly the contour integral, which is more complex than the original eigenvalue problem itself.
To circumvent these two issues, the complex contour integral is approximated with an $N_\text{p}$-point numerical quadrature.
With this approximation, the elements of $S_M$ are expressed as

\begin{equation}
  \mathcal{P}_{M} \Phi^{(i)}_{\text{guess}}
    \approx \frac{1}{2\pi \mathrm{i} } \sum_{k=1}^{N} w_k (z_k  - \mathcal{H})^{-1} \Phi^{(i)}_{\text{guess}} 
      = \sum_{k=1}^{M} w_k \Phi^{(i,k)} \, .
  \label{eq:ProjectorQuadrature}
\end{equation}

The quadrature weight $w_k$ and the complex energy shift $z_k$ for a given quadrature node $k$ are determined by the numerical integration algorithm and, therefore, do not need to be set externally.
The wave function $\Phi^{(i,k)}$ associated with a given quadrature node $k$ and guess state $i$ is obtained by solving the linear system

\begin{equation}
  (z_k - \mathcal{H}) \Phi^{(i,k)} = \Phi^{(i)}_{\text{guess}}
  \label{eq:linfeast}
\end{equation}
that is defined in terms of the same shift-and-invert operator $\Gamma_{z_k} = (z_k - \mathcal{H})$ as in vDMRG[IPI].
The Hamiltonian $\mathcal{H}$ can then be diagonalized within the $S_M$ subspace to obtain the approximate eigenpairs within the energy interval $I_E$.

\begin{figure}[htbp!]
  \centerline{\includegraphics[width=0.4\textwidth]{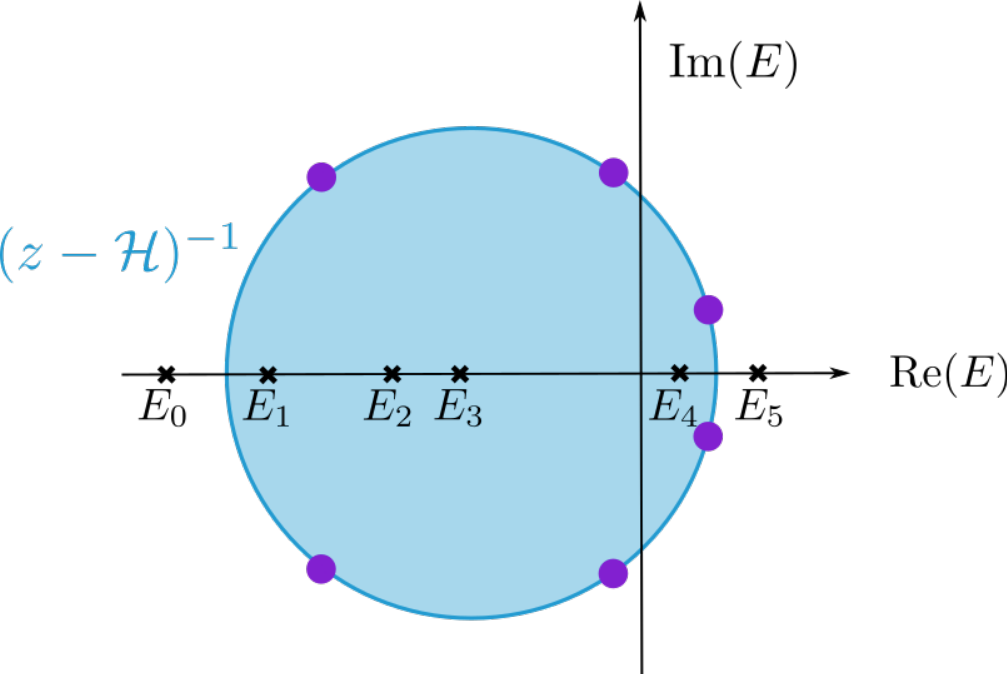}}
  \caption{Graphical representation of the FEAST algorithm.
           The blue circle represents the contour integral needed to calculate the projection operator $(z-\mathcal{H})^{-1}$ in the complex plane.
           The enclosed energy interval $I_E$ contains four eigenvalues, namely $\lbrace E_1, E_2, E_3, E_4 \rbrace$. 
           In this example, the contour integral (blue circle) is approximated by a six-point Gauss-Hermite integration, as depicted by the purple quadrature points.}
  \label{fig:feast}
\end{figure}

It should be noted that if Eq.~(\ref{eq:linfeast}) is solved exactly, the only approximation introduced by the FEAST algorithm is the numerical integration in Eq.~(\ref{eq:ProjectorQuadrature}).
Hence, a single FEAST iteration will return the eigenpairs to arbitrary accuracy for a large enough quadrature grid.
This is a major advantage compared to vDMRG[IPI], where the linear system must be solved $N_{\text{IPI}}$ times in a sequential fashion.
In vDMRG[FEAST], the $M \times N_\text{p}$ linear systems of Eq.~(\eqref{eq:linfeast}) are mutually independent and can be solved in parallel.
By virtue of the trivial parallelization, FEAST is an efficient algorithm to compute multiple excited states simultaneously.
vDMRG[FEAST] can also be applied to energy regions with a high density of states, as it does not require energy guesses for the shift parameter $\omega$ but the entire energy range can instead be calculated at once.

\subsection{Example - Vibrational Transition Energies of Ethlyene}

\begin{figure}[htbp!]
  \centering
  \includegraphics[width=0.7\textwidth]{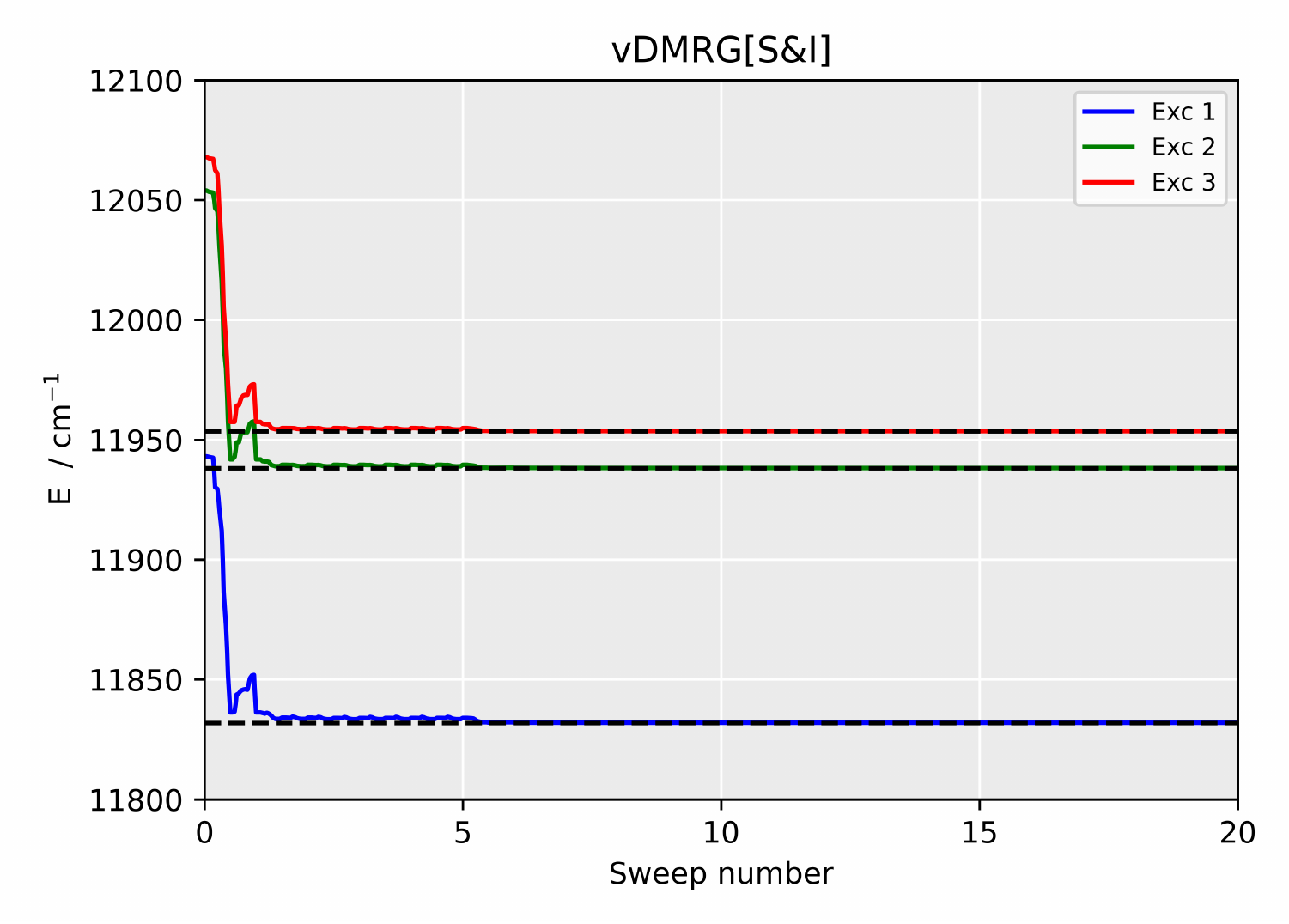}
  \includegraphics[width=0.7\textwidth]{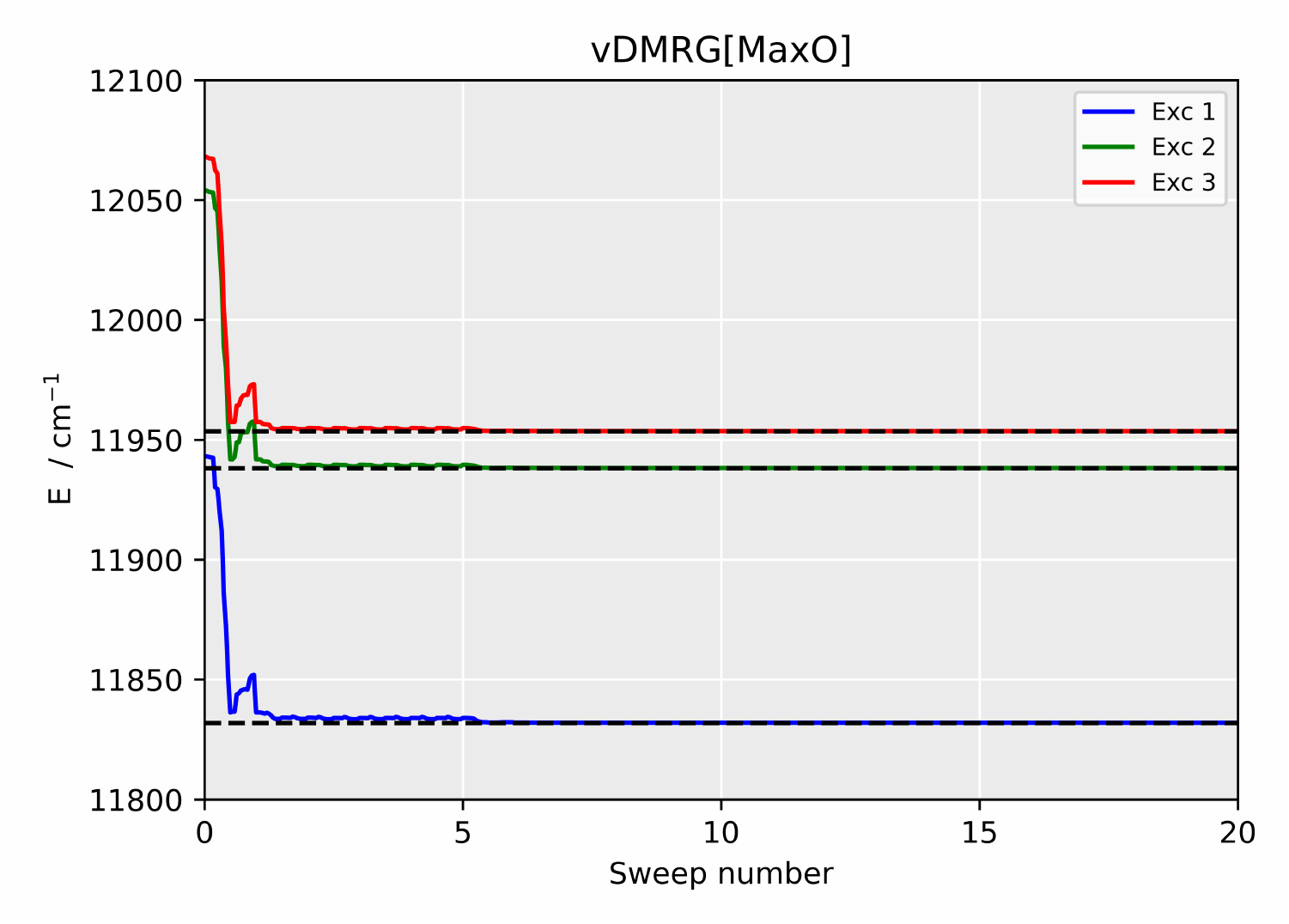}
  \caption{vDMRG[S\&I] (upper panel) and vDMRG[MaxO] (lower panel) energy convergence as applied to the optimization of the first (blue line), second (green line), and third (red line) vibrationally excited state of ethylene.
  We set the shift parameter $\omega$ for vDMRG[S\&I] to 11800~cm$^{-1}$, 11900~cm$^{-1}$, and 11950~cm$^{-1}$ for the first, second, and third excited state, respectively.
  The reference MPS for vDMRG[MaxO] is constructed from the ONV associated with the target vibrational level.
  In both panels, we report the converged energies obtained with vDMRG[ortho] as dashed black lines.
  The bond dimension $m$ was set to 20 in all calculations.}
  \label{fig:ExcitedStates_Ethylene}
\end{figure}

With the excited-state DMRG methods introduced above we calculated the three lowest vibrational excitation energies of ethylene, the same system that we studied in Section~\ref{sec:ex_gs}.
As we show in the upper panel Figure~\ref{fig:ExcitedStates_Ethylene}, the vDMRG[S\&I] optimization converges within 2 sweeps for all $\omega$ values, therefore as fast as for ground state DMRG.
The converged energy of all states matches the reference data obtained with vDMRG[ortho] (represented as dashed black lines in Figure~\ref{fig:ExcitedStates_Ethylene}).
The same trend is observed for vDMRG[MaxO].
In this case, we constructed the reference MPS for the overlap calculation from the ONV associated with the target state.
As we show in the lower panel of Figure~\ref{fig:ExcitedStates_Ethylene}, the vDMRG[MaxO] convergence rate is as fast as for vDMRG[S\&I].
Hence, both the maximum overlap criterion and the shift-and-invert transformation, if taken separately, are sufficient to guide the optimization towards the excited states targeted in this example.
The optimized transition frequencies for the three vibrations are 820.16~cm$^{-1}$, 926.46~cm$^{-1}$, and 941.83~cm$^{-1}$.
As we showed in Ref.~\citenum{baiardi19c}, both of these optimization algorithms become less stable when applied to high-energy states.
The vDMRG[f] algorithm, possibly combined with vDMRG[MaxO], yields the highest numerical stability for these more challenging cases.

\section{Nuclear Dynamics with Matrix Product States}

Tensor network-based methods are not limited to time-independent calculations, but can be applied to any quantum many-body problem.
A particularly interesting application is the solution of the time-dependent Schr\"{o}dinger equation that reads, in Hartree atomic units,

\begin{equation}
  \mathcal{H} \vert \Psi \rangle = \mathrm{i} \frac{\partial \vert \Psi \rangle}{\partial t} \, .
  \label{eq:TDSchro}
\end{equation}

Methods aiming at solving Eq.~(\ref{eq:TDSchro}) by encoding $| \Psi \rangle$ as an MPS are broadly referred to time-dependent DMRG (TD-DMRG) algorithms.\cite{Paeckel2019_Review}
Various TD-DMRG approaches, which have been proposed in the literature, differ in the computational strategy to solve Eq.~(\ref{eq:TDSchro}), but they all share a common limitation, \textit{i.e.}, they all represent a time evolving many-body wave function as MPS throughout the whole propagation.
Whereas the accuracy of time-independent DMRG (TI-DMRG) under certain conditions is guaranteed by the area law,\cite{Hastings2007_AreaLaw} there is no such time-dependent equivalent which would guarantee that a TD wave function can be encoded as a compact MPS.
On the contrary, it has been observed that the wave function entanglement often increases with time under non-equilibrium conditions, a phenomenon known as the ``entanglement barrier effect''.
To further complicate matters, if the TD-DMRG wave function is not accurately represented as an MPS at a given time $t$, this inaccuracy may compromise the simulation accuracy for all subsequent times.

While these observations have questioned the reliability of TD-DMRG, it still remains a promising avenue for efficient quantum dynamics simulations for two reasons.
First, even though the TD-DMRG efficiency cannot be guaranteed \textit{a priori}, the area law does not apply to time-independent vibrational structure problems either due to the presence of long-range many-body interaction terms.
Nevertheless, DMRG was shown to produce sufficiently accurate results for large-scale full CI calculations.
Hence, designing quantum-chemical TD-DMRG algorithms is a key step to assess if and how the entanglement barrier effect impacts the simulation accuracy in practice.
Second, techniques that are already exploited in modern quantum-dynamics methods can be leveraged to tame the entanglement-barrier effect.
For instance, orbital optimization techniques enable adapting the one-particle basis on-the-fly to the non-equilibrium condition and can, therefore, enhance the TD-DMRG efficiency.\cite{Meyer2012_MCTDH-Review,Kurashige2018_MPS-MCTDH,Legeza2019_OrbitalOptimization-TD} \\

In the following, we will first describe the most popular TD-DMRG methods, with particular emphasis on the tangent-space TD-DMRG theory introduced in Refs.~\citenum{Lubich2014_TimeIntegrationTT} and \citenum{Haegeman2016_MPO-TDDMRG} as it is the most promising one in quantum chemical applications.
We will follow the derivation that we reported in our previous works on vibrational\cite{baiardi19b} and electronic\cite{Baiardi2021_ElectronDynamics} dynamics with TD-DMRG.
We will then discuss possible strategies that can be adopted to enhance the TD-DMRG efficiency and tame the entanglement barrier effect.

\subsection{Entanglement Barrier Effect in TD-DMRG}

To illustrate the origin of the entanglement barrier effect, we approximate the solution of Eq.~(\ref{eq:TDSchro}) for a time step $\Delta t$ with a first-order Taylor expansion:

\begin{equation}
  | \Psi(t + \Delta t) \rangle = | \Psi(t) \rangle - \mathrm{i} \Delta t \mathcal{H} | \Psi(t) \rangle
  \label{eq:TDDMRG_FirstOrder}
\end{equation}

If we now encode $| \Psi(t) \rangle$ as an MPS (see Eq.~(\ref{eq:MPS_2})) with time-dependent entries $M_{a_{i-1},a_i}^{\sigma_i}(t)$, and $\mathcal{H}$ as an MPO with entries $\mathcal{H}_{b_{i-1},b_i}^{\sigma_i,\sigma_i'}$ that we assume, for simplicity, to be time independent, then $\mathcal{H} | \Psi(t) \rangle$, can be encoded as follows:

\begin{eqnarray}
  \mathcal{H} | \Psi(t) \rangle &=& \sum_{b_1,\ldots,b_{L-1}} \sum_{a_1,\ldots,a_{L-1}} 
                                   \sum_{\boldsymbol{\sigma}, \boldsymbol{\sigma}', \boldsymbol{\sigma}''}
    H_{1,b_1}^{\sigma_1,\sigma_1'} \cdots H_{b_{L-1},1}^{\sigma_L,\sigma_L'}
    M_{1,a_1}^{\sigma_1''} \cdots M_{a_{L-1},1}^{\sigma_L''}
    | \boldsymbol{\sigma} \rangle \langle \boldsymbol{\sigma}' | \boldsymbol{\sigma}'' \rangle
  \label{eq:MPS_MPO_Contraction_1} \\
    &=& \sum_{b_1,\ldots,b_{L-1}} \sum_{a_1,\ldots,a_{L-1}} 
                                 \sum_{\boldsymbol{\sigma}, \boldsymbol{\sigma}'}
    H_{1,b_1}^{\sigma_1,\sigma_1'} \cdots H_{b_{L-1},1}^{\sigma_L,\sigma_L'}
    M_{1,a_1}^{\sigma_1'} \cdots M_{a_{L-1},1}^{\sigma_L'}
    | \boldsymbol{\sigma} \rangle 
  \label{eq:MPS_MPO_Contraction_2} \\
    &=& \sum_{\substack{a_1,\ldots,a_{L-1} \\ b_1,\ldots,b_{L-1}}} \sum_{\boldsymbol{\sigma}}
    \underbrace{\left( \sum_{\sigma_1'} H_{1,b_1}^{\sigma_1,\sigma_1'} M_{1, a_1}^{\sigma_1'} \right)}_{N_{1, b_1a_1}^{\sigma_1}} 
     \cdots 
    \underbrace{\left( \sum_{\sigma_1'} H_{b_{L-1},1}^{\sigma_L,\sigma_L'} M_{a_{L-1},1}^{\sigma_L'} \right)}_{N_{ b_{L-1}a_{L-1},1}^{\sigma_L}}
    | \boldsymbol{\sigma} \rangle 
  \label{eq:MPS_MPO_Contraction_3} \\
\end{eqnarray}

where $N_{ b_{i-1}a_{i-1},b_i a_i}^{\sigma_i}$ are the tensor entries of the resulting MPS.
The MPS bond dimension variation between sites $i$ and ($i$+1) is $b_i$ times larger than that of the original MPS, which gives rise to the entanglement barrier.
We note, however, that the $b_i$ factor is only an upper bound to the bond dimension increase.
The MPS compression algorithms introduced above can be exploited to reduce, if possible, the bond dimension.
Moreover, we note that the bond dimension growth will increase with $b_i$, \textit{i.e.}, with the MPO bond dimension, which increases with the Hamiltonian complexity and, more specifically, with the extent of the long-range many-body correlations.
Therefore, finding compact MPO representations is crucial to tame the entanglement barrier effect.

\subsection{State-of-the-art TD-DMRG Approaches}

TD-DMRG is maximally efficient for short-ranged Hamiltonians that can be expressed as:

\begin{equation}
  \mathcal{H} = \sum_{i=1}^L h_{i,i+1}
  \label{eq:ShortRangedHamiltonian}
\end{equation}
where $h_{i,i+1}$ is an operator coupling only neighboring sites $i$ and ($i$+1 of the DMRG lattice.
In these cases, instead of approximating the propagation with a first-order Taylor series expansion, the propagator $e^{-\mathrm{i}\mathcal{H}t}$ can be factorized based on the first-order Trotter approximation as:

\begin{equation}
  e^{-\mathrm{i}t\mathcal{H}} \approx \prod_{i \, \text{even}} e^{-i h_{i,i+1} t}
                                      \prod_{i \, \text{odd}} e^{-i h_{i,i+1} t} 
  \label{eq:Trotter}
\end{equation}
and, subsequently, encoded as an MPO with bond dimension $N_i$ for site $i$, $N_i$ being the dimension of the local basis for site $i$.
The resulting propagation scheme, known as the time-evolving block decimation (TEBD) algorithm,\cite{Vidal2004_TEBD} is the most efficient approach to propagate short-range Hamiltonians.
TEBD variants that can be applied to long-range Hamiltonians have also been proposed,\cite{White2010_MinimallyEntangledThermalStates,Bauernfeind2019_Comparison} but they require applying so-called swap gates to bring non-neighboring interacting sites close to one another.
This often leads to a drastic growth of the bond dimension $m$ for complex Hamiltonians, with a consequent increase of the computational cost.

Methods that support arbitrarily complex MPOs are better suited for implementations supporting long-ranged Hamiltonians, and can be divided in two classes.
The first one includes so-called ``propagate-and-compress'' approaches\cite{Shuai2019_TDDMRG-GPU} that solve the TD Schr\"{o}dinger equation by applying conventional integration algorithms, such as the Lanczos\cite{Frahm2019_TD-DMRG_Ultrafast} or the Runge-Kutta\cite{Ronca2017_TDDMRG-Targeting,Ren2018_TDDMRG-Temperature} propagators, by encoding all the basis vectors as MPSs.
Directly solving Eq.~(\ref{eq:TDDMRG_FirstOrder}) belongs to this class of methods, as this corresponds to a straightforward Euler integration of the differential equation.
The common limitation of all these methods is the need for calculating terms as $\mathcal{H}^n | \Psi_\text{MPS} \rangle$, where the accuracy of the integration algorithm grows with $n$.
The resulting wave functions are encoded by highly non-compact MPSs and, therefore, the MPSs must be compressed after each time step to keep the bond dimension $m$ fixed.

As an alternative, more appealing schemes based on the time-dependent variational principle have been developed.
Instead of directly solving Eq.~(\ref{eq:TDSchro}), these methods minimize the Dirac-Frenkel functional $F(t)$, in each time step, defined as:\cite{Moccia1973_TDVP}

\begin{equation}
  F(\Psi(t)) = \left\| \mathcal{H} | \Psi(t) \rangle - \mathrm{i} \frac{ \partial \Psi(t)}{\partial t} \right\|^2 \, ,
  \label{eq:DiracFrenkel}
\end{equation}

At each time $t$, $| \Psi(t) \rangle$ is expressed as an MPS with bond dimension $m$ and the functional $F(t)$ is minimized with respect to the MPS entries to return the best wave function approximation.
The first TD-DMRG algorithm relying on Eq.~(\ref{eq:DiracFrenkel}) was introduced by Haegeman et al.\cite{Haegeman2011_TimeDependentVariationalPrinciple} and propagates all the MPS entries simultaneously for a given time step.
Unfortunately, the equations of motion resulting from such an approach require the inversion of potentially ill-conditioned singular matrices and the algorithm is therefore prone to numerical instabilities.

\subsection{Tangent-space TD-DMRG}

The minimization problem of Eq.~(\ref{eq:DiracFrenkel}) is conveniently expressed as the following projected TD Schr\"{o}dinger equation:\cite{Lubich2014_TimeIntegrationTT,Haegeman2016_MPO-TDDMRG}

\begin{equation}
  \mathcal{P}_{\rm MPS(m)} \mathcal{H} | \Psi(t) \rangle = \mathrm{i} \frac{ \partial \Psi(t)}{\partial t} \, ,
  \label{eq:ProjectedTD}
\end{equation}
with $\mathcal{P}_{\rm MPS(m)}$ being the tangent-space projector onto the manifold composed by all MPSs with bond dimension $m$.
In practice, the tangent-space projector implicitly compresses $\mathcal{H} | \Psi(t) \rangle$ so that it has the same bond dimension as the right-hand side MPS, and therefore, Eq.~(\ref{eq:ProjectedTD}) can be solved exactly, without explicitly compressing the MPS at each time step.
The first key advantage of methods relying on Eq.~(\ref{eq:ProjectedTD}) is that the closed-form expression for $\mathcal{P}_{\rm MPS(m)}$ reads as\cite{Lubich2014_TimeIntegrationTT}

\begin{equation}
  \mathcal{P}_{\rm MPS(m)}
    = \sum_{i=1}^{L}  \sum_{a_i^{(\mathcal{L})} \sigma_{i+1} a_{i+1}^{(\mathcal{R})}} 
      | a_i^{(\mathcal{L})} \sigma_{i+1} a_{i+1}^{(\mathcal{R})} \rangle 
      \langle a_i^{(\mathcal{L})} \sigma_{i+1} a_{i+1}^{(\mathcal{R})} |
    - \sum_{i=1}^{L-1} \sum_{a_{i+1}^{(\mathcal{L})} a_{i+1}^{(\mathcal{R})}} 
      | a_{i+1}^{(\mathcal{L})} a_{i+1}^{(\mathcal{R})} \rangle 
      \langle a_{i+1}^{(\mathcal{L})} a_{i+1}^{(\mathcal{R})} | \, .
  \label{eq:TangentSpaceProjector}
\end{equation}

Eq.~(\ref{eq:TangentSpaceProjector}) defines the $\mathcal{P}_{\rm MPS(m)}$ in terms of the so-called left- and right-renormalized bases.
The left-renormalized bases for site $i$, $\vert a_{i}^{(\mathcal{L})} \rangle$, is defined as follows:

\begin{equation}
  \vert a_{i}^{(\mathcal{L})} \rangle 
    = \sum_{a_{i-1} \sigma_i} M_{a_{i-1},a_i}^{\sigma_i} \vert a_{i-1}^{(\mathcal{L})} \sigma_i \rangle
  \label{eq:LeftRenormalizedBasis}
\end{equation}

The right-renormalized basis is defined analogously as:

\begin{equation}
  \vert a_{i-1}^{(\mathcal{R})} \rangle 
    = \sum_{a_i \sigma_i} M_{a_{i-1},a_i}^{\sigma_i} \vert a_i^{(\mathcal{R})} \sigma_i \rangle
  \label{eq:RightRenormalizedBasis}
\end{equation}

The left and right renormalized basis for site $i$ are defined with the MPS canonized with respect to the same site, and therefore, each term of Eq.~(\ref{eq:TangentSpaceProjector}) is defined based on a different MPS canonization.
This is, however, not a problem because Eq.~(\ref{eq:ProjectedTD}) can be solved based on the Trotter approximation: the terms of the sum appearing in Eq.~(\ref{eq:TangentSpaceProjector}) are applied subsequently, for increasing $i$ values, by adapting the MPS canonization consequently.
Specifically, the first differential equation to be solved reads:

\begin{equation}
  \sum_{a_0^{(\mathcal{L})} \sigma_0 a_1^{(\mathcal{R})}} 
  | a_0^{(\mathcal{L})} \sigma_{0} a_{1}^{(\mathcal{R})} \rangle \langle a_0^{(\mathcal{L})} \sigma_{0} a_{1}^{(\mathcal{R})} | \mathcal{H} | \Psi(t) \rangle 
    = \mathrm{i} \frac{ \partial \Psi(t)}{\partial t} \, ,
  \label{eq:ProjectedTD_1}
\end{equation}
which is solved for a time-step $\Delta t$.
The resulting solution defines the initial wave function for the second differential equation, that reads

\begin{equation}
  - \sum_{a_1^{(\mathcal{L})} \sigma_0 a_1^{(\mathcal{R})}} 
    | a_1^{(\mathcal{L})} a_{1}^{(\mathcal{R})} \rangle \langle a_1^{(\mathcal{L})} a_{1}^{(\mathcal{R})} | \mathcal{H} | \Psi(t) \rangle 
      = \mathrm{i} \frac{ \partial \Psi(t)}{\partial t} \, .
  \label{eq:ProjectedTD_2}
\end{equation}

Eqs.(\ref{eq:ProjectedTD_1}) and (\ref{eq:ProjectedTD_2}) are solved for every site on the DMRG lattice to obtain the wave function at time $t$.
Hence, for each time step the lattice is traversed in a sweeping fashion, as in TI-DMRG.
A key advantage of tangent-space-based methods is that Eqs.~(\ref{eq:ProjectedTD_1}) and (\ref{eq:ProjectedTD_2}) are linear differential equations with constant coefficients that can be solved to arbitrary high accuracy with Krylov-based methods.\cite{Saad1992_MatrixExponential}

\begin{figure}[htbp!]
  \centering
  \includegraphics[width=.75\textwidth]{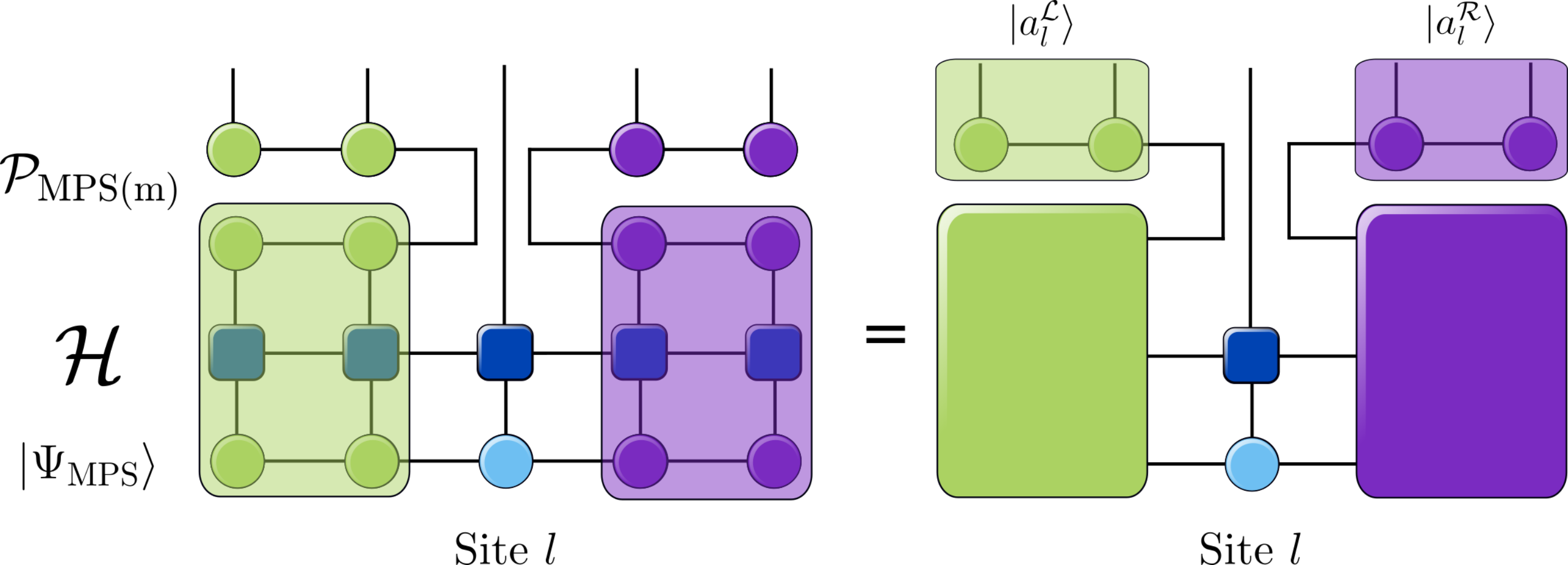}
  \caption{Graphical representation of the tensor network to be contracted in order to evaluate Eq.~(\ref{eq:ProjectedTD_1}) for the site $l$ of the DMRG lattice.}
  \label{fig:TangentSpace}
\end{figure}

As we show in Figure~\ref{fig:TangentSpace}, the left-hand side of Eq.~(\ref{eq:ProjectedTD_1}) (and its generalization to any site $i$ of the DMRG lattice) can be expressed in terms of the left and right boundaries that we already introduced for time-independent DMRG.
Therefore, the solution of Eq.~(\ref{eq:ProjectedTD_1}) reads:

\begin{equation}
  \textbf{M}^{\sigma_0}(t+\Delta t) = e^{-\mathrm{i}\Delta t \textbf{H}_0^{(1)}} \textbf{M}^{\sigma_0}(t) \, ,
  \label{eq:SolutionProjectedTD_1}
\end{equation}
with $\textbf{H}_0^{(1)}$ being the local Hamiltonian representation that is diagonalized in conventional DMRG for site 0.
Note that the graphical representation of Figure~\ref{fig:TangentSpace} holds for MPS canonized with respect to the same site at which the tangent-space projector is centered.
This is, however, not a limitation, since the MPS normalization can be moved along the lattice while the terms of Eq.~(\ref{eq:TangentSpaceProjector}) are applied, as is done in TI-DMRG.
The possibility of implementing the tangent-space TD-DMRG method with algorithms that are very similar to those of TI-DMRG is the key advantage of this method.

By applying the Trotter factorization, propagating the whole MPS for a given time step $\Delta t$ requires traversing the DMRG lattice to solve Eqs.~(\ref{eq:ProjectedTD_1}) and (\ref{eq:ProjectedTD_2}) for all sites of the DMRG lattice.
At the end of the propagation, the MPS is canonized with respect to the last site.
The canonization should therefore be shifted to the first site before applying the propagator for the next time step.
However, as suggested in Ref.~\citenum{Haegeman2016_MPO-TDDMRG}, the propagators can be applied in the reversed order instead, analogously to a backward sweep in conventional, time-independent DMRG.
This approach corresponds to a second-order Trotter approximation of the solution to Eq.~(\ref{eq:ProjectedTD}) and, therefore, enhances the method's accuracy and avoids unnecessary shifting of the MPS canonization.

As we already highlighted above, the main challenge associated with TD-DMRG simulations is the fact that the wave function entanglement evolves with time and the bond dimension $m$ should be adapted correspondingly to yield a constant wave function accuracy over the whole propagation.
However, the projection operator defined in Eq.~(\ref{eq:TangentSpaceProjector}) constrains the bond dimension $m$ to be the same over the whole propagation.
Such an approach does not allow for a dynamical adaptation of $m$.
This means, in practice, that if the wave function entanglement is maximum at a given time $t$, then the bond dimension $m$ yielding an accurate wave function representation at that time must be used for the whole propagation in order to obtain converged results.
Moreover, the bond dimension $m$ is constrained to be the same as that of the wave function at $t$=0, \textit{i.e.} at the beginning of the propagation.
The algorithm fails therefore in describing the dynamical increase of the wave function entanglement. \\

All these limitations are addressed by the two-site TD-DMRG variant that generalizes the definition of Eq.~(\ref{eq:TangentSpaceProjector}) as follows:\cite{Haegeman2016_MPO-TDDMRG}

\begin{equation}
  \mathcal{P}_{\rm MPS(m)}^\text{TS}
    = \sum_{i=1}^{L}  \sum_{\substack{a_i^{(\mathcal{L})} a_{i+2}^{(\mathcal{R})} \\ \sigma_i \sigma_{i+1}}}
      | a_i^{(\mathcal{L})} \sigma_{i} \sigma_{i+1} a_{i+2}^{(\mathcal{R})} \rangle 
      \langle a_i^{(\mathcal{L})} \sigma_{i} \sigma_{i+1} a_{i+2}^{(\mathcal{R})} |
    - \sum_{i=1}^{L-1} \sum_{\substack{a_i^{(L)} a_{i+1}^{(\mathcal{L})} \\ \sigma_i}}
      | a_{i+1}^{(\mathcal{L})} a_{i+1}^{(R)} \rangle \langle a_{i+1}^{(\mathcal{L})} a_{i+1}^{(R)} |
  \label{eq:TangentSpaceProjectorTS}
\end{equation}

The differential equation associated with Eq.~(\ref{eq:TangentSpaceProjectorTS}) can be solved based on the Trotter approximation as introduced above.
Note that the propagation associated with the first term of Eq.~(\ref{eq:TangentSpaceProjectorTS}) will change the bond dimension $m$.
The MPS must therefore be compressed to keep the bond dimension fixed, in the spirit of time-independent two-site DMRG.
As noted in Ref.~\citenum{Baiardi2021_ElectronDynamics}, this two-site variant is largely more accurate in the presence of rapidly varying time-dependent perturbations that induce a corresponding fast variation of the wave function entanglement.

\subsection{Quantum chemical applications of real-time TD-DMRG}

As any quantum dynamics method, TD-DMRG has two primary application fields: the calculation of molecular spectra and the simulation of ultrafast molecular processes. \\
An absorption spectrum $I(\omega)$ can be expressed, within the dipole approximation and at T=0~K (the inclusion of thermal effects will be discussed in the next section), as the integral over the dipole autocorrelation function $C_{\mu}(t)$ with

\begin{equation}
  I(\omega)  = \alpha \int_{-\infty}^{+\infty} \, C_{\mu}(t) \, e^{\mathrm{i}\omega t} \, \text{d} t
  = \alpha \int_{-\infty}^{+\infty} \, 
    \langle \Psi | \mu e^{-\mathrm{i} \mathcal{H} t} \mu | \Psi \rangle \, e^{\mathrm{i}\omega t} \, 
    \text{d} t \, ,
  \label{eq:AbsorptionSpectrum-TD}
\end{equation}
with $| \Psi \rangle$ being the ground-state wave function and $\alpha$ a proportionality constant.
The definition of the dipole operator $\mu$ depends on the simulation target.
For vibrational spectroscopies, $\mu$ contains the nuclear contribution to the molcular dipole moment while, for electronic spectroscopies, it is the electric dipole operator.\cite{Neville2018_TD-Electronic,Baiardi2021_ElectronDynamics}
Eq.~(\ref{eq:AbsorptionSpectrum-TD}) can be calculated with TD-DMRG by encoding $\mu | \Psi \rangle$ as an MPS and propagating the resulting wave function under the action of the Hamiltonian $\mathcal{H}$.\cite{baiardi19c}
This procedure enables calculating spectra without resorting to complex sum-over-states expressions that are obtained with frequency-domain approaches.
The time-domain route to molecular spectroscopy has been applied to TD-DMRG for calculating absorption spectra of molecular aggregates\cite{Ren2018_TDDMRG-Temperature,Kloss2019_Multiset-MPS,Shuai2019_TDDMRG-GPU} based on excitonic Hamiltonians.
The development of the tangent-space TD-DMRG methods has opened the route towards extending this scheme to more complex \textit{ab initio} vibronic Hamiltonians as well.\cite{Kurashige2018_MPS-MCTDH,baiardi19c}

Eq.~(\ref{eq:AbsorptionSpectrum-TD}) expresses a spectrum as the Fourier transform of the dipole autocorrelation function $C_{\mu}(t)$, and the integral over the whole propagation time averages out the details of the dynamics.
On the one hand, the $C_{\mu}(t)$ function decays for $t \rightarrow +\infty$ due to dephasing and decoherence effects.
Therefore, the long-time dynamics, which are hard to capture with TD-DMRG due to the entanglement barrier effect, have only a minor impact on the calculation of $I(\omega)$ and spectra converge remarkably fast with the bond dimension, as we showed in Ref.~\citenum{baiardi19c}.
On the other hand, a TD-DMRG propagation contains much more information than that exploited by Eq.~(\ref{eq:AbsorptionSpectrum-TD}).
For instance, the availability of the many-body wave function for all times $t$ makes it possible to fully characterize the driving forces of the non-equilibrium quantum dynamics.
For instance, when applied to exciton dynamics, TD-DMRG does not only yield spectra, but can also identify the pathways that are followed by excitons to propagate along a molecular aggregate.\cite{Borrelli2017_ExcitonTemperature,baiardi19c,Kloss2019_Multiset-MPS}

\subsection{Example - Absorption Spectrum of Pyrazine}

We demonstrate the efficiency of the tangent-space TD-DMRG method on the simulation of the vibrationally-resolved absorption spectrum of pyrazine.
The presented data is taken from our work on the vibrational TD-DMRG theory,\cite{baiardi19b} where we applied TD-DMRG on the 4-mode vibronic Hamiltonian defined in Ref.~\citenum{Raab1999_Pyrazine}, which is one of the most common benchmark systems for molecular quantum dynamics algorithms.
This vibronic Hamiltonian describes the coupled S$_1$ and S$_2$ electronically excited states as:

\begin{equation}
   \mathcal{H} = \left(
      {\begin{array}{cc}
    	\mathcal{T}_{S_1}(\bm{Q}) & 0 \\
    	0 & \mathcal{T}_{S_2}(\bm{Q}) \\
       \end{array} } \right) + \left(
      {\begin{array}{cc} 
    	\mathcal{V}_{S_1}(\bm{Q}) & \mathcal{V}_{12}(\bm{Q}) \\
    	\mathcal{V}_{12}(\bm{Q})  & \mathcal{V}_{S_2}(\bm{Q}) \\
       \end{array} }
    \right)
  \label{eq:PES_Pyrazine}
\end{equation}
where $\mathcal{V}_{S_1}(\bm{Q})$ and $\mathcal{V}_{S_2}(\bm{Q})$ are the diabatic PES for the S$_1$ and S$_2$ states, respectively, and $\mathcal{V}_{12}(\bm{Q})$ is the non-adiabatic coupling between them.
Both the PES and the non-adiabatic coupling are approximated as a second-order Taylor expansion in terms of the vibrational modes $\bm{Q}$.
By applying Eq.~(\ref{eq:AbsorptionSpectrum-TD}) to the vibronic Hamiltonian defined in Eq.~(\ref{eq:PES_Pyrazine}), we simulated in in Ref.~\citenum{Raab1999_Pyrazine} the vibronic spectrum of pyrazine for the S$_1$ $\leftarrow$ S$_0$ and S$_2$ $\leftarrow$ S$_0$ excitations.

\begin{figure}[htbp!]
  \centering
  \includegraphics[width=.49\textwidth]{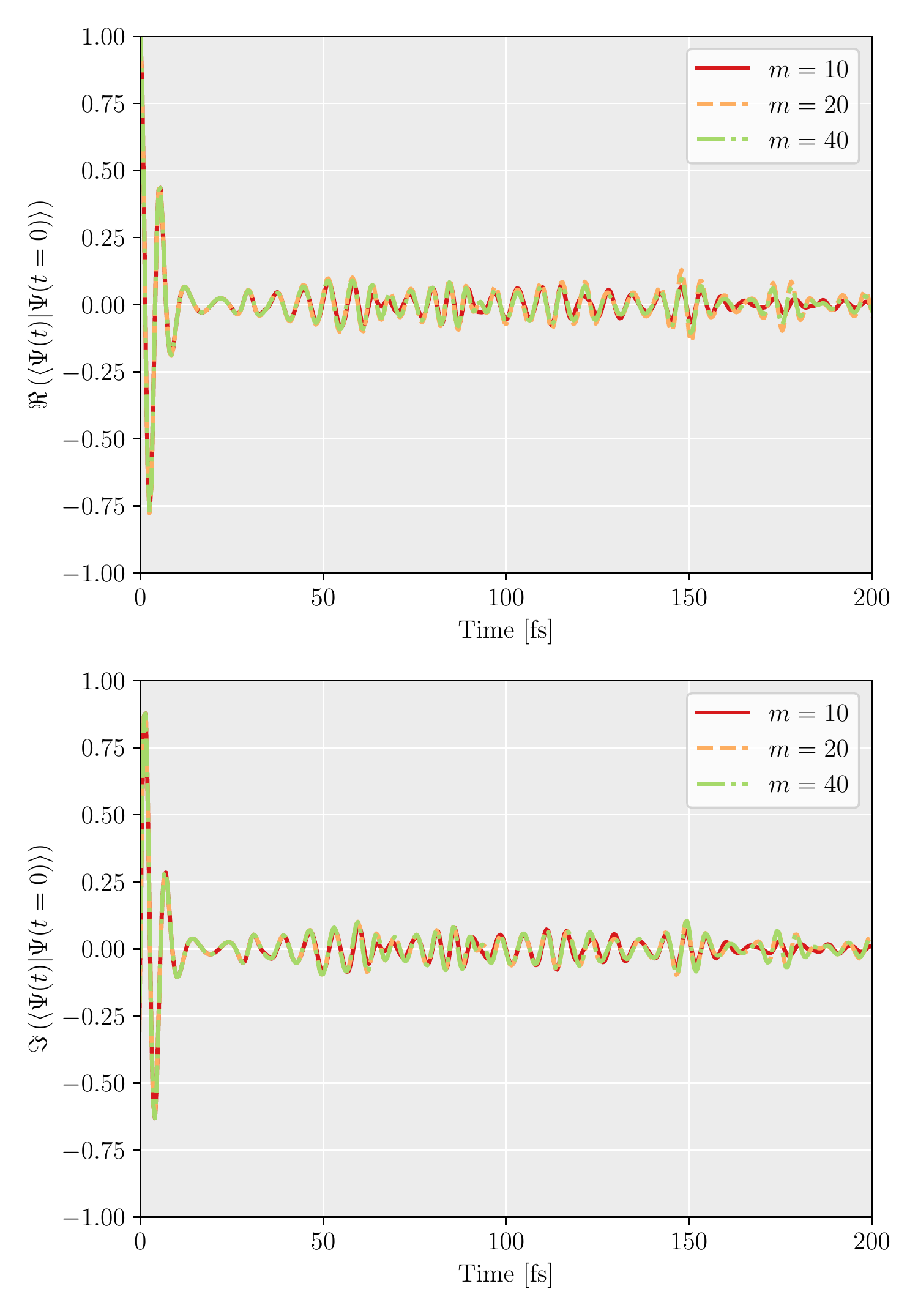}
  \includegraphics[width=.49\textwidth]{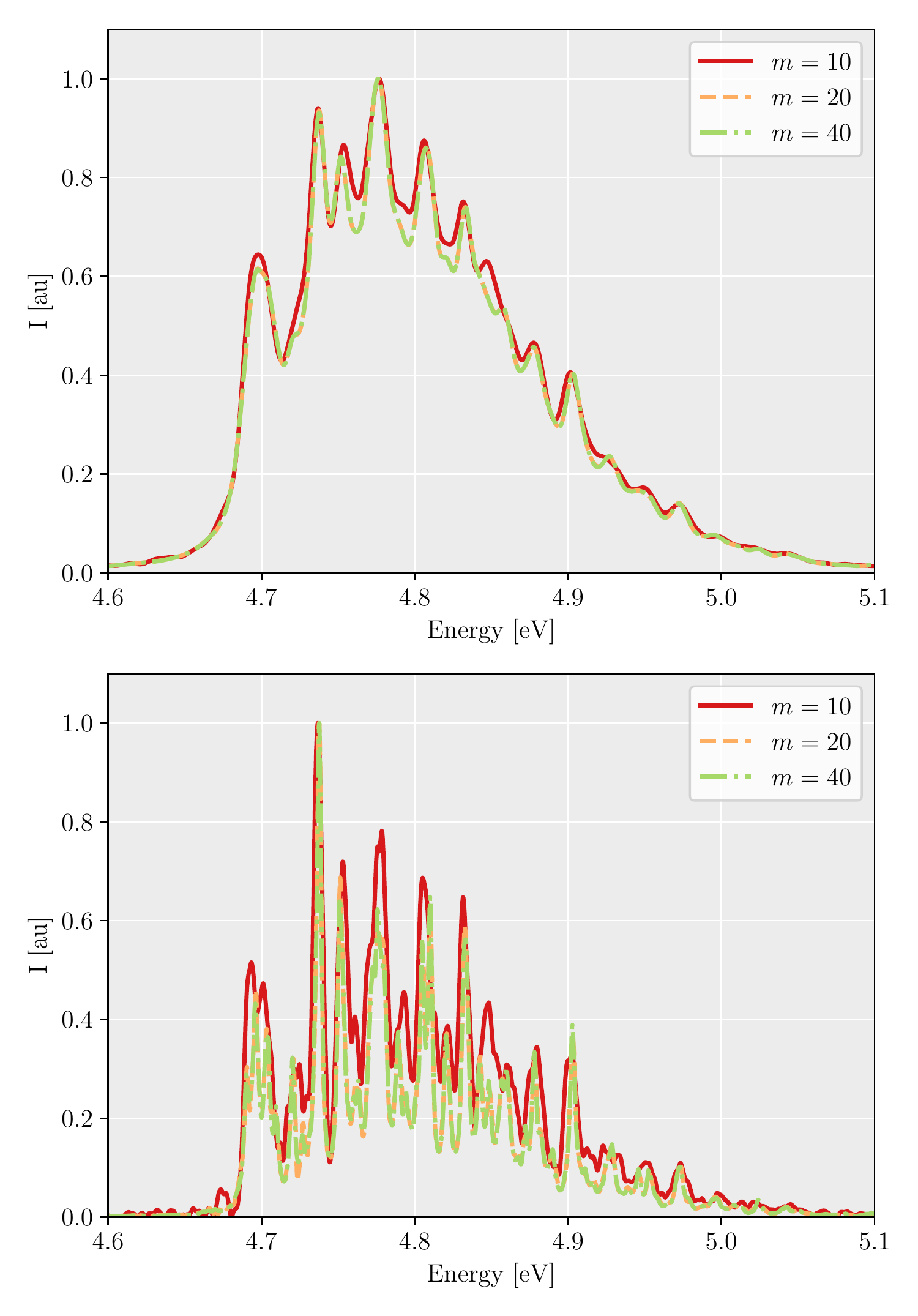}
  \caption{Real (upper left panel) and imaginary (lower left panel) part of the time-dependent dipole autocorrelation function $C_\mu(t)$ calculated based on the 4-mode vibronic Hamiltonian of pyrazine defined in Eq.~(\ref{eq:PES_Pyrazine}).
  The corresponding spectrum obtained with Gaussian broadening functions with a half-width at half-maximum of 200~cm$^{-1}$ (upper panel) and 50~cm$^{-1}$ (lower panel) are reported in the right part of the figure.
  TD-DMRG calculations were carried out with $m$=10 (solid red line), $m$=20 (dashed orange line) and $m$=40 (dashed green line).
  The time-step $\Delta t$ is 0.5~fs in all cases.
  This figure has been generated with data taken from Ref.~\citenum{baiardi19b}.}
  \label{fig:Spectra}
\end{figure}

We report in the left part of Figure~\ref{fig:Spectra} the time-dependent autocorrelation function $C_\mu(t)$, taken from the data reported in our original TD-DMRG work.\cite{baiardi19b}
$C_\mu(t)$ was qualitatively converged already with a bond dimension of $m$=10 and quantitative differences compared to the $m$=20 results were observed only starting from 100~fs.
No changes were observed by further increasing $m$ to 40.
As expected, the absorption spectrum, obtained from the Fourier transformation of the autocorrelation function and reported in the right panel of Figure~\ref{fig:Spectra}, followed the same convergence pattern.
For the higher-resolution spectrum, obtained with a HWHM of 50~cm$^{-1}$, small differences were observed between the intensity of the individual vibronic bands calculated with $m$=10 and $m$=20.
Nevertheless, the band shape was fully converged with $m$=40.
By including broadening effects with Gaussian functions with a half-width at half-maximum (HWHM) of 200~cm$^{-1}$, the band shape was converged already with $m$=10.
As we noted in Ref.~\citenum{baiardi19b}, the convergence of the spectrum with respect to $m$ did not change substantially for larger vibronic Hamiltonians.
The absorption spectrum calculated based on the full-dimensional 24-mode Hamiltonian of pyrazine converged with a low $m$ value of 100.
This suggests that the entanglement barrier effect impacts, in practice, only marginally the simulation accuracy.
Hence, TD-DMRG is a powerful method for nuclear quantum dynamics simulations.

\subsection{Imaginary-time TD-DMRG: ground-state optimization and thermal ensembles}

The tangent-space TD-DMRG method can be extended to complex-valued times $t$ and applied to two classes of problems of special interest, namely to purely imaginary-time propagation and to time values that contain both a real and an imaginary part.
The propagation algorithm obtained for purely imaginary time values $t$ reads:

\begin{equation}
  \frac{\partial | \Psi_\text{MPS}(t) \rangle}{\partial t} = - \mathcal{P}_{\rm MPS(m)} \mathcal{H} | \Psi_\text{MPS} \rangle
  \label{eq:ImaginaryTimePropagation}
\end{equation}
and converges to the ground state for $t \rightarrow +\infty$ independently on the initial wave function, provided that the guess MPS and the final, converged wave function are not orthogonal.
Eq.~(\ref{eq:ImaginaryTimePropagation}) provides the foundation of many modern electronic-structure calculations, such as the full-CI quantum Monte Carlo (FCIQMC)\cite{Alavi2009_FCIQMC} and the auxiliary-field quantum Monte Carlo\cite{Zhang2003_AFQMC} methods.
The imaginary-time variant of MCTDH is also routinely used to optimize anharmonic vibrational wave functions.\cite{Meyer2006_ImaginaryTimeMCTDH,Wang2014_ImaginaryTimeMCTDH,Wodraszka2018_ImaginaryTimeMCTDH}
The imaginary-time TD-DMRG method (iTD-DMRG) is often less efficient than conventional TI-DMRG and requires more sweeps to converge.
However, iTD-DMRG features the key advantage of reliably optimizing the right eigenvalues of non-Hermitian Hamiltonians that are obtained through similarity transformation.\cite{Baiardi2020_tcDMRG}
Consider, for instance, the CI equation associated with Jastrow-like wave function ansatz:

\begin{equation}
  \mathcal{H} e^{\mathcal{J}} | \Psi \rangle = E e^{\mathcal{J}} | \Psi \rangle
  \label{eq:JastrowSchrodinger}
\end{equation}

The Jastrow factor can be revolved into the Hamiltonian operator as follows:

\begin{equation}
  e^{-\mathcal{J}} \mathcal{H} e^{\mathcal{J}} | \Psi \rangle = E | \Psi \rangle
  \label{eq:JastrowSchrodinger2}
\end{equation}
and $| \Psi \rangle$ can be calculated as a right eigenvalue of the non-Hermitian operator $e^{-\mathcal{J}} \mathcal{H} e^{\mathcal{J}}$, usually referred to as the transcorrelated Hamiltonian.\cite{Boys1969_Transcorrelated,Handy1969_Transcorrelated,Baiardi2020_tcDMRG}
For electronic-structure methods, the Jastrow factor can be tuned to include short-range correlation effects, so that $| \Psi \rangle$ can be encoded as a more compact CI wave function.
Such an approach can drastically enhance both the FCIQMC\cite{Alavi2018_Transcorrelated} and the DMRG\cite{Baiardi2020_tcDMRG} convergence.
Even though extensions of such schemes to vibrational-structure problems have been scarce, the generality of the transcorrelated approach makes it appealing also for applications to vibrational-structure calculation methods.

Further generalizing the TD-DMRG algorithm to time values that contain both a real and an imaginary part makes it applicable to the simulation of thermal ensembles.
For instance, the absorption spectrum $I(\omega)$ associated with a given Hamiltonian $\mathcal{H}$ is proportional, under the dipole approximation, to the time-dependent dipole-dipole autocorrelation function $C(t)$ that reads:

\begin{equation}
  C(t) = \langle \Psi |  e^{-\beta \mathcal{H}} e^{\mathrm{i} \mathcal{H} t} \mu 
                         e^{-\mathrm{i} \mathcal{H} t} \mu e^{-\beta \mathcal{H}} | \Psi \rangle
  \label{eq:TD_Absorption_Spectrum}
\end{equation}
with $\beta = (k_B T)^{-1}$, $k_B$ being the Boltzmann constant and $T$ the temperature.
Eq.~(\ref{eq:TD_Absorption_Spectrum}) can be evaluated, within a DMRG-based framework, by encoding $e^{-\beta \mathcal{H}} | \Psi \rangle$ as an MPS obtained by propagating the $T=0$ wave function in the imaginary time for a time equal to $\beta$.
The resulting MPS is then propagated in real time to obtain the correlation function $C(t)$ and, therefore, to calculate the spectral observable.
Such a strategy has been combined with multiple TD-DMRG variants,\cite{Karrash2012_FiniteTemperature-SpinChain,Ren2018_TDDMRG-Temperature} including the tangent-space one,\cite{Shuai2020_TangentSpaceTDDMRG-CarrierMobility} to include temperature effects in non-equilibrium simulations of model excitonic Hamiltonians.
The generality of the MPS/MPO-based framework introduced above will make it possible to simulate also more complex \textit{ab-initio} vibronic Hamiltonians within the very same theoretical framework.

\subsection{Enhancing the TD-DMRG efficiency}

The DMRG algorithm can be interpreted as an efficient algorithm to solve large-scale full-CI problems in a given basis set.
The accuracy of a full CI-based wave function is, however, often not required for many chemical applications.
For instance, MCTDH does not represent the wave function as a full-CI expansion in the primitive basis (often obtained as distributed Gaussian\cite{Hamilton1986_DistributedGaussians} or discrete variable representation\cite{Colbert1992_DVR} functions).
Instead, it contracts the primitive basis to obtain a pre-optimized modal basis and, then, applies full-CI only to a subset of the resulting modal basis in the spirit of active space-based methods.
The same idea can be exploited in tensor-based methods as well.
Kurashige combined in Ref.~\citenum{Kurashige2018_MPS-MCTDH} the vibrational DMRG theory with the MCTDH algorithm to propagate both the single-particle functions and the MPS simultaneously.
Therefore, the MPS is encoded as

\begin{equation}
  | \Psi_\text{MPS}(t) \rangle = \sum_{n_1 \cdots n_L} \sum_{a_1, \ldots, a_{L-1}}
    M_{1,a_1}^{n_1}(t) M_{a_1,a_2}^{n_2}(t) \cdots M_{a_{L-1},1}^{n_L}(t) 
    \prod_{i=1}^{L} \chi_{k_i}^i (Q_i, t) \, ,
  \label{eq:MPS_MCTDH}
\end{equation}
$\chi^i_{k_i}(Q_i, t)$ being the $k_i$-th single-particle function for mode $i$.
The main challenge in deriving the equation of motion associated with the \textit{ansatz} of Eq.~(\ref{eq:MPS_MCTDH}) is that the modal variation that is parallel to the modals that are included in Eq.~(\ref{eq:MPS_MCTDH}) is already included in the propagation of the CI coefficients, expressed as MPSs in Eq.~(\ref{eq:MPS_MCTDH}).
To avoid double-counting, the equation-of-motion for the modals $\chi^i_{k_i}(Q_i, t)$ should only contain contributions arising from the modals that are not included in Eq.~(\ref{eq:MPS_MCTDH}).
Ref.~\citenum{Kurashige2018_MPS-MCTDH} avoids this problem by applying an algorithm originally designed for MCTDH\cite{Bonfanti2018_MCTDH-ProjectorSplitting} that relies on the so-called single-hole function formalism.

The optimization scheme introduced above relies on the partition of the modal basis into an ``active'' and a ``virtual'' set, and DMRG is applied only to the former modals.
However, Legeza and co-workers\cite{Legeza2019_OrbitalOptimization-TD} introduced algorithms to further optimize the active modal basis to yield the most compact MPS wave function at all times.
Originally introduced for time-independent problems,\cite{Legeza2016_OrbitalOptimization} this algorithm relies on the two-site TD-DMRG variant.
After the two-site tensor is propagated, and before it is compressed to yield the target bond dimension $m$, the single-particle functions associated with the optimization sites are rotated based on a unitary transformation that is optimized to minimize the pair entropy.
In this way, the wave function entanglement and, therefore, the bond dimension can be effectively reduced without compromising the accuracy.
The success of this orbital optimization scheme as applied to electron dynamics suggests that it can also enhance the efficiency of nuclear dynamics simulations.
We note, however, that the algorithm introduced in Ref.~\citenum{Legeza2016_OrbitalOptimization} cannot be straightforwardly applied to nuclear dynamics because electrons are indistinguishable fermionic particles, while vibrational modes are described in terms of distinguishable, bosonic particles, as discussed above.
Within the $n$-mode second quantization framework, the unitary transformation can only be applied between pairs of sites associated with the same mode.
Otherwise, basis functions associated with different particles would be mixed.
Each site is associated with a different mode within the canonical quantization framework, and this prevents a straightforward application of the algorithm of Ref.~\citenum{Legeza2016_OrbitalOptimization} to the corresponding vDMRG variant.
The pair entropy minimization principle cannot be applied in this case but the modals can still be optimized separately for each individual site as suggested in Ref.~\citenum{Jeckelmann2015_BosonicOptimization}.

\section{Conclusion and Outlook}

We introduced the vibrational Density Matrix Renormalization Group (vDMRG) method\cite{baiardi17,baiardi19c} to calculate vibrational ground- and excited-states of molecular systems.
Even if the DMRG algorithm has originally been designed to target short-range spin Hamiltonians, we found\cite{baiardi17} it to be, out-of-the-box, as efficient as alternative state-of-the-art methods when applied to vibrational-structure calculations.
This shows that tensor network based methods are a promising approach to overcome the current limits of large-scale vibrational structure calculations.

However, many research routes remain to be explored to further enhance vDMRG.
Following approaches devised for electronic-structure calculations,\cite{Zgid2008_DMRG-SCF,Ma2017_SecondOrder-SCF} vDMRG can be combined with modal optimization schemes to further augment the representation power of the vibrational tensor network states for both time-independent\cite{Rauhut2010_V-MCSCF} and time-dependent\cite{Meyer2012_MCTDH-Review} simulations.
Moreover, vDMRG has, so far, only been applied to solve the vibrational full CI problem.
The high accuracy of vDMRG is, however, only essential to describe strongly anharmonic molecular vibrations, while faster methods based on perturbation theory\cite{Barone2005_VPT2,Rauhut2010_VMP2} are sufficient for an accurate description of weakly anharmonic modes.
Hybrid variational-perturbative schemes are currently the method of choice to efficiently describe strongly-correlated electronic wave functions.\cite{Morokuma2013_CASPT2,Freitag2017_DMRG-NEVPT2}
Extending this combined approach to vibrational problems would make it possible to efficiently target molecules with several dozen degrees of freedom.

Another, more radical improvement of the vDMRG algorithm could be achieved by generalizing it beyond the matrix product state parameterization.
As we highlighted in this chapter, the key component of vDMRG is the availability of an efficient optimization scheme that is based on the alternating least-squares algorithm.
This optimization procedure cannot be extended to multi-dimensional tensor network states, with few exceptions,\cite{verstraete04,Legeza2015_TTNS,Olivares2015_DMRGInPractice} and this limitation has impeded the development of fully automated multidimensional tensor network-based methods so far.
vDMRG has been applied mostly to moderately anharmonic potentials and, in these cases, an MPS yields a very compact representation of the CI wave function.
However, for strongly anharmonic molecules,\cite{McCoy2004_CH5,Bowman2017_WaterCluster} more complex tensor network states will provide a more flexible representation of the wave function and are, therefore, an appealing alternative to MPS-based DMRG.

A key assumption underlying the theoretical framework described in this chapter is the Born--Oppenheimer (BO) approximation that simplifies the molecular Hamiltonian by disentangling the nuclear and electronic motions.
Perturbation theory offers an effective way to account for the effects that are neglected by BO--based calculations,\cite{Pachucki2008_NonAdiabatic,Pachucki2009_NonAdiabatic} such as non-adiabatic couplings or quantum electrodynamics effects.
The generality of the MPS/MPO-based framework makes it possible to straightforwardly include these terms in the definition of the vibrational Hamiltonian.
However, perturbative approaches rely on BO-based calculations that require calculating the PES, a task that may become challenging for strongly-anharmonic molecules.
Pre-Born--Oppenheimer (PreBO) methods overcome this problem by solving the full molecular Schr\"{o}dinger equation by treating nuclear and electronic degrees of freedom on the same footing.
Explicitly-correlated algorithms are the method of choice for accurate PreBO calculations on small molecules,\cite{Matyus2012_MolecularStructure,Bubin2013_Review} but their computational cost makes them applicable to few-particle molecules only.
Orbital-based methods\cite{Pavosevic2020_PreBO-Review} have a much smaller computational cost, but they struggle with capturing strong nuclear-electron correlation effects and, therefore, do not yield spectroscopically accurate simulations.
Tensor network states are inherently designed to capture strong correlation effects and are, therefore, the ideal candidate to fill the gap between explicitly correlated and orbital-based methods.
Various PreBO DMRG variants are currently under development.\cite{White2019_PreBO,Muolo2020_NEAP-DMRG}
Although the development of these methods is still in its infancy, the possibility of encoding efficiently strong correlation effects in DMRG may make orbital-based PreBO methods amenable to accurate vibrational-structure calculations.

\providecommand{\latin}[1]{#1}
\makeatletter
\providecommand{\doi}
  {\begingroup\let\do\@makeother\dospecials
  \catcode`\{=1 \catcode`\}=2 \doi@aux}
\providecommand{\doi@aux}[1]{\endgroup\texttt{#1}}
\makeatother
\providecommand*\mcitethebibliography{\thebibliography}
\csname @ifundefined\endcsname{endmcitethebibliography}
  {\let\endmcitethebibliography\endthebibliography}{}

\end{document}